\documentclass[12pt]{article}
\usepackage{amsmath}

\usepackage{enumitem}
\usepackage{natbib}
\usepackage{amssymb} 
\usepackage{tikz} 
\usepackage{float} 
\usetikzlibrary{decorations.pathreplacing} 
\usepackage[colorlinks=true, allcolors=blue!80]{hyperref}

\title{\textbf{Conglomerates, Liquidity Shocks, and\\ Innovation-Led Growth}\\\vspace{0.5em}
\large Working Paper – May 2025}

\author{
  Payne Hennigan\thanks{School of Economics, Peking University. Email: \texttt{paynehennigan@gmail.com}} 
}

\date{\today}

\begin{document}

\maketitle
\vspace{-1.5em} 
\begin{abstract}
\noindent
I develop a dynamic model of how internal capital markets in conglomerates respond to liquidity shocks when affiliated firms vary in innovation potential. A two-stage framework defines cutoff rules for when the conglomerate should liquidate low-productivity firms, coerce intermediate types into short-termist strategies, or preserve high-potential firms for long-horizon R\&D. Embedding these margins into an endogenous growth model, I show how the optimal policy evolves: early in development, coercion preserves liquidity while sustaining broad innovation; as the economy nears the frontier and short-term returns decline, the optimal strategy shifts toward binary reallocation between liquidation and long-termism. I characterize two policy failures: a “coercion trap,” where short-termism persists too long, and a “liquidation fallacy,” where viable firms are discarded prematurely. The framework provides microfoundations for dynamic reallocation in conglomerate systems and offers policy insights for crisis-era restructuring.

\vspace{1em}
\noindent\textit{JEL Classification:} H54, E62, H63 \\
\noindent\textit{Keywords:} Endogenous Growth, Conglomerates, Credit Shocks, Internal Capital Markets, Innovation Strategy
\end{abstract}
\vspace{2em}


\section{Introduction}

When firms inside a conglomerate face financial stress, who gets saved and who gets cut? The answer can shape not only short-term survival but also long-term innovation. Conglomerates routinely shift capital across their divisions, deciding whether to redirect funds toward short-term cash generators or preserve long-horizon R\&D projects. These internal decisions can determine which technologies get developed and which are abandoned. In bank-based economies such as Japan, Korea, Germany, and China, conglomerates and business groups have long played a central role in allocating capital across affiliated firms.\footnote{See \citet{Branstetter2002HasJI} and \citet{heshmati2011chaebol} on internal capital markets in East Asia; and \citet{Deeg2001InstitutionalCA} on the German case.} But this mode of centralized control is no longer limited to these settings. In the U.S. and Europe, private equity firms, holding companies, and vertically integrated multinationals increasingly manage liquidity and investment decisions across their portfolios.\footnote{See \citet{Bernstein2022TheEO, Kaplan2008LeveragedBA} for the role of private equity and \citet{Bernstein2019PEfragile} on managing liquidity within business groups. See also \citet{harrigan2024rise} on the ``new conglomerates'' in the digital age.}

The central question is how conglomerates should respond to financial stress when affiliated firms differ in both their near-term liquidity contribution and long-term productivity value. In the face of a liquidity shock, the conglomerate can exert pressure on firms to change strategy. The conglomerate faces a trade-off between two adjustment margins. On one hand, it can \textit{coerce} firms—forcing them to pursue lower-risk, short-term strategies that generate immediate cash flow but reduce long-run output. On the other hand, it can \textit{liquidate} firms—fully withdrawing support and reallocating funds to more promising divisions. After an investment shock, the conglomerate chooses which firms to coerce, liquidate, or preserve. I formalize this in a two-stage model in which each firm selects between two distinct innovation strategies. One is short-term and imitative, generating front-loaded returns with limited lifetime value. The other is long-term and novel, producing higher future gains but weaker liquidity potential in the short-term.\footnote{I refer to ‘short-term’ strategies broadly as innovation modes that yield early returns but low dynamic spillovers—this includes imitation, incremental improvement, and exploitative innovation. In contrast, ‘long-term’ strategies involve high-upside but delayed-payoff innovation, such as frontier R\&D or exploratory research.} I derive closed-form cutoff rules that characterize the optimal allocation. These margins affect not only immediate liquidity, but also the long-term innovation capacity of the group.

These static rules become dynamic margins when embedded in an endogenous growth model. A key (though standard) assumption is that both types of innovation become more difficult as the economy approaches the global technology frontier, but the relative benefits of short-termism deteriorate more rapidly \citep{Acemoglu2006DistanceTF}. This shifts the relative payoff of each strategy over time, and hence the optimal cutoff rules derived earlier. In early stages, a wide coercion margin can efficiently preserve internal liquidity while maintaining broad innovative capacity, since the benefits of catchup are relatively large, while the front-loaded payoffs act as a source of liquidity. In later stages, the coercion margin shrinks, and it becomes the better option to transition to a binary strategy without coercion where the worse performers are sacrificed to maintain liquidity for the group and the better performers are preserved for long-term investment.  I show, furthermore, how failure to adapt cutoffs in response to this changing payoff landscape leads to long-run inefficiency and stagnation. I examine two forms of dynamic misallocation. The coercion trap arises when the conglomerate continues to coerce too many firms into short-run strategies even as the economy approaches the frontier, failing to reoptimize its cutoff policies. The liquidation fallacy reflects the opposite bias: an overemphasis on liquidation to preserve long-term potential, leading to underutilization of viable short-run options. Both behaviors generate persistent distortions in capital allocation and delay convergence to optimal growth.

In this paper I make three primary contributions. First, I develop a theoretical framework in which a conglomerate reallocates internal capital across affiliated firms pursuing heterogeneous innovation strategies—some focused on short-term imitation, others on long-horizon R\&D. While internal capital markets have been widely studied in corporate finance (e.g., \citet{scharfstein2000theds}; \citet{Rajan2000CostDiv}; \citet{Goplan2014InternalCapDiv}), existing theory emphasizes agency frictions, soft-budget constraints, or governance inefficiencies, and does not address how conglomerates allocate capital across divisions with different innovation horizons. At the same time, empirical work shows that internal capital markets influence firm-level R\&D investment (e.g., \citet{Lamont1997CashFA}; \citet{shin1998intcapeff}; \citet{belenzon2019innbusgroup}), yet there is no theory linking this to optimal reallocation under financial stress. This paper fills that gap by modeling a conglomerate as a centralized allocator facing a real-side liquidity constraint and internal trade-offs between immediate cash flow and future innovation. The framework endogenizes the decision to liquidate, coerce into short-termism, or preserve long-run R\&D—linking liquidity management directly to innovation composition.

Second, I embed this mechanism into an endogenous growth model in which the conglomerate’s capital reallocation—via liquidation, coercion, or preservation—determines the economy’s growth path. While canonical endogenous growth models emphasize innovation heterogeneity and creative destruction (e.g., \citet{aghion1992model}; \citet{acemoglu2018micro}), they abstract from the institutional structure governing capital allocation. I fill this gap by introducing a centralized allocator that adjusts reallocation margins in response to liquidity shocks. This links theories of misallocation under financial frictions (e.g., \citet{buera2011financial}) with corporate finance models of internal capital markets, and complements macro-financial work on liquidity-constrained intermediaries (e.g., \citet{gertler2010financial}; \citet{gertler2011model}) and growth under financial constraints (e.g., \citet{aghion2010volatility}). Unlike those settings, where firms face external borrowing limits (or the financial intermediary itself faces a constraint on lending), I internalize liquidity allocation within a hierarchical firm group, offering a new channel through which institutional structure shapes aggregate growth. 

Third, I analyze how the optimal internal reallocation policy should evolve with development as the economy approaches the global technology frontier. In early stages, short-term (imitative) strategies are effective and liquidity-conserving. But as the return to imitation declines, long-run innovation becomes relatively more valuable, and reallocation policy must adapt. I highlight how rigid allocation rules to manage liquidity shocks can lead to persistent misallocation if not adjusted over time. Two stylized scenarios illustrate this point: a coercion trap, where all firms are preserved but pushed into increasingly ineffective short-termism, and a liquidation fallacy, where no firms are saved but prematurely discarded despite long-run potential. These distortions arise not from standard borrowing frictions, but from institutional inertia. This complements existing work on misallocation in growth models (e.g., \citet{akcigit2017growth}) and innovation across development (e.g., \citet{Acemoglu2006DistanceTF}), while echoing the insight of \citet{dewatripont1995credit} that static policies can underperform when institutional responses lag behind a changing economic environment. This is not a generic call for market-based allocation. The conglomerate structure remains intact throughout. The key result is that inefficiencies arise not from internal capital markets per se, but from their failure to adjust reallocation policy as the economy develops.

The paper proceeds as follows. Section~\ref{sec:lit_review} gives a summary of the literature. Section~\ref{sec:model} introduces the two-period allocation model and derives the cutoff conditions. Section~\ref{sec:growth} embeds the model into a dynamic framework and analyzes the endogenous evolution of strategy. Section~\ref{sec:policy} discusses dynamic misallocation, institutional rigidity, and policy implications. Appendices provide full derivations for all results.


\section{Background and Related Literature} \label{sec:lit_review}

This paper builds on several strands of literature spanning corporate finance, macroeconomic growth, and the theory of internal capital markets. It contributes to the broader understanding of how large business groups allocate capital in the face of liquidity shocks, with implications for innovation and long-run growth.

I build first on the corporate finance literature that examines how internal capital markets shape investment under financial constraints. Foundational work such as \citet{rajan1992insiders} and \citet{vonthadden1995long} analyzes how internal finance can mitigate external market frictions by reallocating liquidity within organizations, especially when projects have multi-period investment needs. \citet{rajan1992insiders} emphasizes how insiders allocate capital more efficiently in the presence of asymmetric information, while \citet{vonthadden1995long} highlights internal agency costs in long-term investment. Later work focuses more explicitly on capital allocation across divisions.  \citet{stein1997internal} models how headquarters may distort divisional investment, and \citet{scharfstein2000theds} show how divisional rent-seeking can lead to inefficient resource allocation. \citet{Rajan2000CostDiv} further connects internal misallocation to the diversification discount, suggesting inefficiencies grow with organizational complexity. Empirical studies such as \citet{Lamont1997CashFA}, \citet{shin1998intcapeff}, and \citet{belenzon2019innbusgroup} document that internal capital markets affect investment and R\&D, particularly during periods of liquidity stress. These papers underscore the importance of internal finance for investment efficiency, but rarely model how capital is reallocated within firm groups in response to evolving liquidity constraints or long-horizon innovation incentives. 

My focus is not on agency frictions or rent-seeking, but on how internal allocation rules interact with liquidity shocks and development-stage innovation trade-offs—highlighting how static internal policies can generate persistent misallocation even in the absence of standard market failures. I treat conglomerates not just as substitutes for external markets, but as evolving allocators whose effectiveness depends on the adaptability of internal cutoff rules. Internal capital markets, in this view, do not belong exclusively to either bank-based or market-based systems—as traditionally framed in the literature (e.g., \citet{allen2000comparing}, \citet{demirguckunt2001bankmarket})—but can perform both functions dynamically. 

A large literature in endogenous growth theory explores how innovation drives long-run productivity, emphasizing forward-looking investment under uncertainty. Foundational models such as \citet{romer1990endogenous}, \citet{aghion1992model}, and \citet{Acemoglu2006DistanceTF} formalize innovation as a cumulative process requiring upfront capital and delivering delayed gains. More recent work introduces firm heterogeneity and explores how financing shocks affect capital reallocation and innovation dynamics, including \citet{bianchi2019growth} and \citet{guerronquintana2019financial}. Other models investigate how crises interact with innovation timing and recovery speeds, such as \citet{barlevy2004timing}, \citet{anzoategui2019endogenous}, and \citet{queralto2020slow}. Studies like \citet{bonciani2023slow} and \citet{cerra2021financial} further connect innovation slumps to financial downturns. Despite these advances, this literature typically employs representative firms or abstract frictions, and does not model within-group capital allocation or reallocation strategies across heterogeneous innovation types. The internal organizational response to liquidity stress—particularly in firm groups or conglomerates—remains relatively underexplored. I extend this literature by introducing a microfounded model of conglomerate-level reallocation across heterogeneous firms pursuing either short- or long-term innovation, and embed it in an endogenous growth environment where optimal reallocation policy evolves with development.

A complementary body of research explores how growth dynamics evolve as economies approach the global technology frontier. In canonical models of endogenous growth, such as \citet{aghion1992model} and \citet{Acemoglu2006DistanceTF}, imitation and innovation coexist, with the balance between them shifting over time. These models suggest that catch-up strategies based on short-term learning can be effective during early development but diminish in productivity as the frontier nears. I extend this insight by modeling internal capital reallocation within conglomerates as development-dependent: early-stage environments favor liquidity-conserving strategies (e.g., coercion to sustain low-productivity firms), while mature economies benefit from riskier, long-horizon investments. Static allocation rules, however, may persist beyond their optimal window, producing endogenous misallocation. This mechanism resonates with findings in \citet{akcigit2017growth}, which document how policy and institutional inertia can delay the transition to innovation-led growth.

I contribute as well to a broader macroeconomic literature on structural change in development, which examines how economies transition from capital-intensive production toward knowledge-driven, innovation-led growth. Foundational models such as \citet{acemoglu1997prometheus} and \citet{buera2011financial} emphasize how financial frictions can distort this process by delaying reallocation toward high-productivity sectors. A general result across this literature is that imitation-based strategies and intermediary-led finance are more effective in early development, when the distance to the global frontier is large. As economies mature and invention becomes more central, market-based finance becomes more efficient due to its superior ability to allocate capital across decentralized, innovation-driven projects \citep{lin2022distance}. Empirical work by \citet{demirguckunt2001bankmarket} and theoretical comparisons by \citet{allen2000comparing} suggest that the optimal financial structure evolves with development. In early stages, bank-based systems dominate due to their superior monitoring and intertemporal risk-sharing capacity, making them especially effective for imitative or capital-intensive investments. As economies approach the global frontier and innovation becomes more decentralized and risky, market-based finance gains importance, offering better cross-sectional risk-sharing and accommodating a diversity of investor beliefs. This reinforces the importance of adapting financial institutions—including internal capital markets—to the changing innovation landscape across development stages. My model adds to this literature by focusing on intra-group allocation dynamics within conglomerates, showing how rigid internal cutoff rules can reinforce these developmental traps even in the absence of external credit constraints. In fact, in my setup internal capital markets are not statically bank-like or market-like, but can dynamically shift along both intertemporal and cross-sectional risk-sharing dimensions depending on the stage of development and institutional adjustment.

I finally contribute to a growing literature on the macroeconomic role of conglomerates and internal capital markets, particularly in East Asia and other emerging economies. While much of the traditional development literature focuses on firms in isolation, work by \citet{almeida2007internal} and  \citet{gopalan2007internal} examines how business groups operate as internal capital markets, partially substituting for weak external finance. My paper complements these studies by developing a theoretical model of group-level reallocation under financial stress and embedding it in an endogenous growth framework, where internal capital markets evolve with development. Empirically, these mechanisms echo the experiences of Japan and Korea, where group-based financial systems emerged to manage long-term investment risk during periods of structural transformation. Studies of Japan’s main bank system (e.g., \citet{horiuchi1995japan}, \citet{wu2012mainbank}) emphasize how centralized monitoring and patient capital helped firms survive through periods of macroeconomic volatility, facilitating gradual sectoral upgrading. In Korea, the 1997 crisis catalyzed sweeping reforms inspired by Washington Consensus prescriptions—prioritizing liberalization, market-based finance, and the dismantling of cross-subsidization within chaebols. However, post-crisis data suggest sluggish credit reallocation and a persistent decline in innovation and productivity growth \citep{queralto2020slow, furman1998eastasia}. These experiences have led many scholars to caution that such reforms, when applied prematurely or without institutional adaptation, can produce lasting developmental traps \citep{aghion2005quality}. My perspective complements work by \citet{hoshi2001corporate}, who show that group finance in Japan and Korea historically responded endogenously to volatility, but also faced declining allocative efficiency as economic conditions changed. The mechanism I propose—misallocation due to institutional inertia within conglomerates—offers a new channel through which structural transformation may stall, even in the absence of binding external financial constraints. In doing so, the paper contributes to broader debates on state intervention, firm boundaries, and financial frictions in development.


\section{The Conglomerate’s Problem}
\label{sec:model}

This section presents the static core of the paper: a two-stage model of capital allocation within a conglomerate-financed system. A representative conglomerate manages a continuum of firms that differ in innovation potential and faces a liquidity shock that limits available investment. In response, it reallocates funding across firms using three margins: liquidation, coercion to short-termism, and preservation for long-termism. These discrete choices generate cutoff rules that later become dynamic policy margins in the endogenous growth model (Section~\ref{sec:growth}).

\subsection{Strategy Payoffs and Assumptions}

The conglomerate begins by drawing a unit mass of firms indexed by innovation type \( \lambda \in [0,1] \). There is no intertemporal persistence in firm identity or state. All firms require the same fixed investment \( I(\lambda) = \bar{I} \), which the conglomerate intends to provide. Assume that the conglomerate has sufficient capital \( L \) to fund all firms at the outset, so that \( L = \int_{0}^{1}I(\lambda)d\lambda\). 

Firms differ in innovation type \( \lambda \in [0,1] \). Each chooses a strategy horizon \( h \in \{s, l\} \) corresponding to a short-term (\( s \)) or long-term (\( l \)) strategy. The model unfolds over two stages. Payoffs over each stage are:
\begin{align}
X_1^h(\lambda) &= \lambda \theta_1^h X_1 \\
X_2^h(\lambda) &= \lambda \theta_2^h X_2
\end{align}
Here, \( \theta_t^h \) indexes strategy-specific productivity, and \( X_t \) are stage-specific scale factors common across firms.

I impose three assumptions:
\begin{enumerate}[label=(\roman*), leftmargin=*]
    \item \( \theta_1^s \geq \theta_1^l \): short-termism generates higher first-stage returns.
    \item \( \theta_2^l > \theta_2^s \): long-termism generates higher second-stage returns.
    \item \( \theta_1^l X_1 + \theta_2^l X_2 > \theta_1^s X_1 + \theta_2^s X_2 \): long-termism yields higher lifetime value.
\end{enumerate}

Absent financial frictions, all firms would pursue the long-term strategy, since it strictly dominates short-termism in lifetime productivity. However, when a liquidity shortfall arises, the conglomerate must reallocate capital by either liquidating some firms entirely or coercing others into short-term strategies that yield earlier returns. Assume that at stage 0 that the size of the liquidity shock is unknown, but that all payoffs are known. There is no possibility of firm failure outside of the conglomerate-wide liquidity shock. The investment size is uncorrelated with unknown firm productivity $\lambda$, so there is no advantage to pre-emptive saving.

\subsection{Liquidity Shock and Reallocation Margins}

At the beginning of the interim stage 1, the conglomerate faces a liquidity shock \( \Delta L \) that reduces available capital to \( L - \Delta L \). This forces the conglomerate to reallocate internally across firms. In order to raise liquidity, the conglomerate must choose between two reallocation margins: coercion and liquidation. I assume that firms must be sustained with $4X_1^l(\lambda)$ in the interim period to deliver final-stage output. This assumption simplifies the dynamic structure and allows for tractable cutoff-based reallocation rules. Without loss of generality, I normalize interim output to be recycled as investment. Each firm must be assigned to one of three categories:
\begin{enumerate}[label=(\roman*), leftmargin=*]
    \item \textit{Liquidation:} Firms are dropped entirely. Investment is canceled and existing firm value is completely liquidated, giving the conglomerate a one-time liquidity gain equal to the interim stage output of the firm:
	\begin{equation}
	\text{Liquidity gained} = X_1^l(\lambda)
	\end{equation}
    \item \textit{Coercion:} Firms are forced into a short-termist strategy that generates greater early returns at the cost of future productivity. Reassigning a firm from long-term to short-term strategy yields a liquidity gain equal to the difference in first-stage output and what is necessary to sustain firm production:
	\begin{equation}
	\text{Liquidity gained} = X_1^s(\lambda) - X_1^l(\lambda)
	\end{equation}
	This captures the substitution from backloaded to frontloaded output while maintaining the necessary level to sustain production over the entire timeline.
    \item \textit{Preservation:} Firms are continued to be fully funded to pursue a long-term strategy that emphasizes second-stage gains.
\end{enumerate}

\subsection{Conglomerate’s Problem}

The conglomerate's choice is governed by two cutoff points: \( \lambda_s \) (the coercion threshold) and \( \lambda_l \) (the preservation threshold). Firms with \( \lambda < \lambda_s \) are liquidated; those with \( \lambda_s \leq \lambda < \lambda_l \) are coerced into short-termism; and those with \( \lambda \geq \lambda_l \) are preserved for a long-termist strategy.

\vspace{1em}
\begin{center}
\begin{tikzpicture}
	\draw[thick] (0,0) -- (10,0);
	\draw[thick] (0,0.3) -- (0,-0.3) node[below] {$0$}; 
	\draw[thick] (3.33,0.3) -- (3.33,-0.3) node[below] {$\lambda_s$};
	\draw[thick] (6.67,0.3) -- (6.67,-0.3) node[below] {$\lambda_l$};
	\draw[thick] (10,0.3) -- (10,-0.3) node[below] {$1$};

	\draw[decorate, decoration={brace, amplitude=12pt}] (0,.5) -- (3.33,.5)
		node[midway, above=12pt] {\small Liquidate};
	\draw[decorate, decoration={brace, amplitude=12pt}] (3.33,.5) -- (6.67,.5)
		node[midway, above=12pt] {\small Coerce (Short-Term)};
	\draw[decorate, decoration={brace, amplitude=12pt}] (6.67,.5) -- (10,.5)
		node[midway, above=12pt] {\small Preserve (Long-Term)};
\end{tikzpicture}
\end{center}
\vspace{-1em}

The conglomerate chooses \( (\lambda_s, \lambda_l) \) to meet the liquidity demands while maximizing final-stage output. Formally, the conglomerate’s problem is:

\begin{equation}
\max_{\lambda_s, \lambda_l} \quad \int_{\lambda_s}^{\lambda_l} X_2^s(\lambda) \, d\lambda + \int_{\lambda_l}^1 X_2^l(\lambda) \, d\lambda
\end{equation}

subject to the liquidity constraint met by the sum of liquidity gained from coercion and liquidation::
\begin{equation}
\Delta L = \int_{\lambda_s}^{\lambda_l} \left[ X_1^s(\lambda) - X_1^l(\lambda) \right] d\lambda + \int_0^{\lambda_s} X_1^l(\lambda) d\lambda.
\end{equation}

\paragraph{Timeline and Decision Sequence}

To give a brief summary, the research process and conglomerate allocation problem unfolds as follows:

\begin{description}[leftmargin=0cm, labelsep=1em, style=standard, font=\bfseries]
    \item[Stage 0:] The conglomerate draws firms indexed by innovation potential \( \lambda \in [0,1] \), each requiring fixed investment \( \bar{I} \). All firms are initially assigned to a long-termist strategy.

    \item[Stage 1:] A liquidity shock \( \Delta L \) occurs, constraining total funding. The conglomerate chooses cutoffs \( (\lambda_s, \lambda_l) \), assigning each firm to liquidation, coercion to short-termism, or preservation for long-termism. Based on the conglomerate decision, firms produce interim  output based on their assigned strategy \( X_1^h(\lambda) \), with \( h \in \{s, l\} \) denoting a short-term or long-term strategy.

    \item[Stage 2:] Firms complete production and deliver second-stage output \( X_2^h(\lambda) \).
\end{description}
The conglomerate’s goal is to maximize long-run output from surviving firms, subject to a liquidity constraint imposed by the shock \( \Delta L \).

\subsection{Interior Solution and Analytical Cutoffs}

When the liquidity shock is moderate, the conglomerate adopts an interior solution. This involves two internal cutoffs, \( \lambda_s \in (0,1) \) and \( \lambda_l \in (0,1) \), with \( \lambda_s \leq \lambda_l \). The conglomerate balances the marginal benefit of preserving long-term productivity (the opportunity cost of coercion) against the marginal liquidity gain from reallocation. The optimal interior cutoffs satisfy the first-order condition:
\begin{equation}
\frac{\theta_2^l - \theta_2^s}{\theta_2^l} = \frac{\theta_1^s - \theta_1^l}{\theta_1^l} \cdot \frac{\lambda_s}{\lambda_l - \lambda_s}.
\label{eq:interior_cutoff_condition}
\end{equation}

Define the ratio of the first-stage productivity loss from coercion to the second-stage productivity gain from liquidation as:
\begin{equation}
    \Gamma \equiv \frac{(\theta_2^l - \theta_2^s)\theta_1^l}{(\theta_1^s - \theta_1^l)\theta_2^l}
\label{eq:Gamma}
\end{equation}

This yields a direct relationship between the cutoffs:
\begin{equation}
    \lambda_l = \frac{1+\Gamma}{\Gamma}\lambda_s
\label{eq:lambda_s}
\end{equation}

For a full derivation, see \ref{app:interior_cutoff}.\\

Using the conglomerate’s liquidity constraint:
\begin{equation}
    \Delta L = \frac{1}{2}X_1\left[(\theta_1^s - \theta_1^l)\lambda_l^2 + (2\theta_1^l - \theta_1^s)\lambda_s^2\right]
\end{equation}

and substituting the relationship between \(\lambda_l\) and \(\lambda_s\), I obtain a single closed-form analytical solution for the liquidation cutoff \(\lambda_s\):
\begin{equation}
\lambda_s = \sqrt{\frac{2\Delta L}{X_1\left[(\theta_1^s - \theta_1^l)(\frac{1+\Gamma}{\Gamma})^2 + (2\theta_1^l - \theta_1^s)\right]}},
\label{eq:lambda_k_closed}
\end{equation}

with \(\Gamma\) as defined above. A full derivation is given in \ref{app:lambda_s}.\\

These expressions show how the conglomerate’s optimal cutoffs depend jointly on first-stage parameters and liquidity needs \((\Delta L, X_1, \theta_1^s, \theta_1^l)\) and second-stage parameters \((\theta_2^s, \theta_2^l)\). The coercion margin, defined by \(\lambda_l\), becomes more attractive as the relative second-stage payoff gap (\(\theta_2^l - \theta_2^s\)) from short-term strategies declines or as the first-stage liquidity benefit (\(\theta_1^s - \theta_1^l\)) from coercion rises. The liquidation margin, \(\lambda_s\), grows when either the liquidity constraint tightens (\(\Delta L\) rises) or when the long-term payoff advantage (\(\theta_2^l - \theta_2^s\)) from preservation weakens.

Figure~\ref{fig:full_solution_set} illustrates the conglomerate’s optimal cutoff strategy under a representative parameterization:

\begin{figure}[H]
    \centering
    \includegraphics[width=0.75\textwidth]{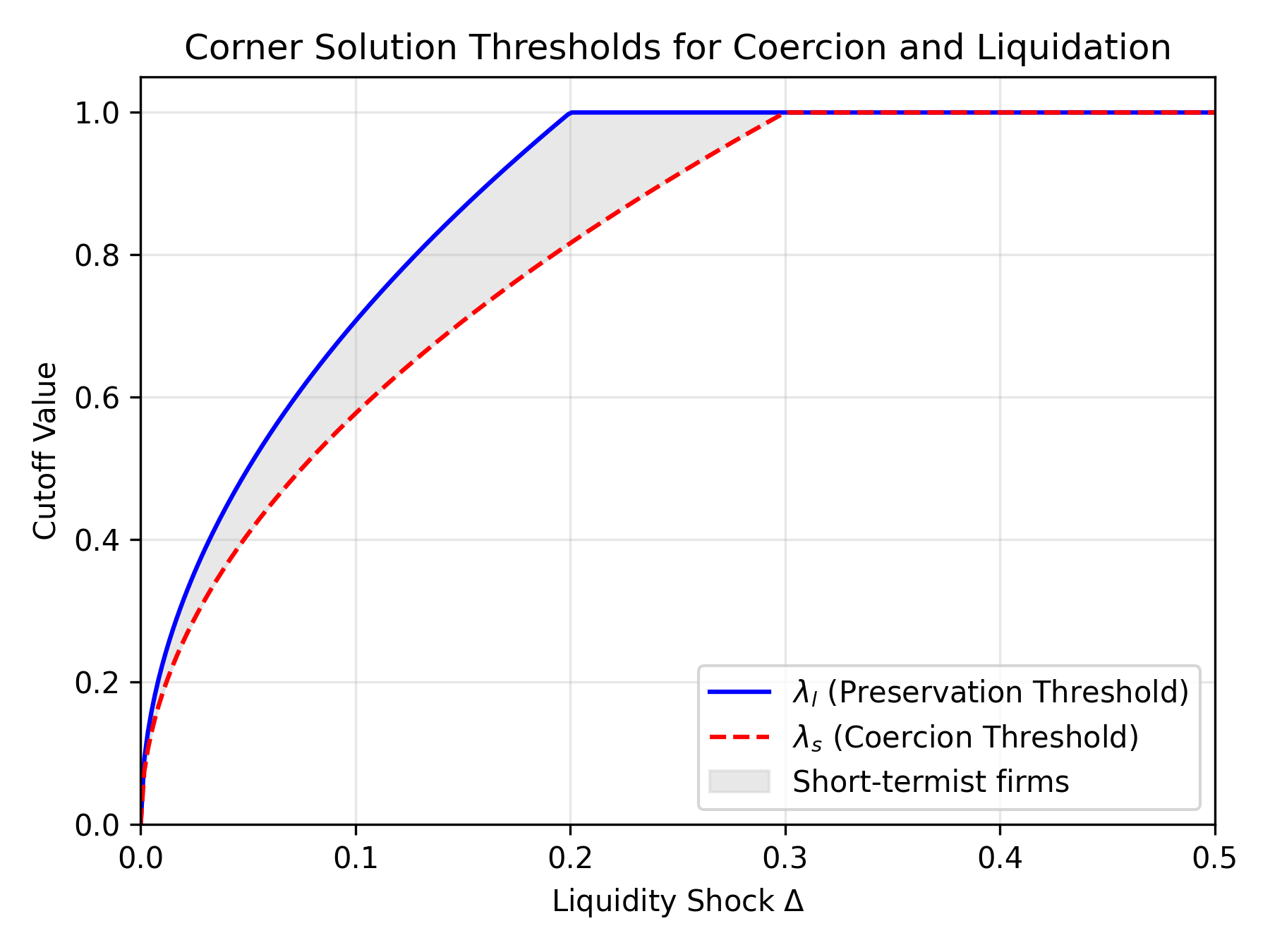}
    \caption{Optimal Cutoff Strategy as a Function of Liquidity Shock. Solid line: preservation threshold \( \lambda_l \); dashed line: coercion threshold \( \lambda_s \). This figure uses a representative parameter set \( (X_1 = 1, \theta_1^l = 0.3, \theta_1^s = 0.7) \) to illustrate how cutoff strategies evolve with increasing liquidity stress. Full derivations confirming the cutoff values and regime transitions are provided in Appendix~A.5.}
    \label{fig:full_solution_set}
\end{figure}

This parameter set satisfies all feasibility conditions in the static model and highlights the transition between regimes. The conglomerate’s cutoff values respond nonlinearly to the size of the liquidity shock \( \Delta L \), and exhibit distinct reallocation strategies depending on whether the interior or corner solution applies. At low levels of \( \Delta L \), the interior solution applies and both margins are used. As \( \Delta L \) approaches 0.2, the coercion cutoff reaches its upper bound \( \lambda_l = 1 \), triggering a transition to a high-shock corner regime with full coercion and partial liquidation. Once \( \Delta L \geq 0.3 \), even coercion and liquidation together are insufficient to fund the required investment—no feasible cutoff configuration exists, and the conglomerate must liquidate all firms. This failure boundary illustrates the limits of internal reallocation under financial stress. For more details see \ref{app:figure}.

These static cutoff regimes provide a foundation for the dynamic analysis in Section~\ref{sec:growth}, where the conglomerate’s problem recurs in each period but the relative payoff to different innovation strategies evolves endogenously as the economy develops. In this extended framework, short-term (imitative) and long-term (novel) innovation become progressively more difficult as the economy approaches the technological frontier, but the liquidity value of short-termism declines faster. This generates a shifting policy landscape in which the optimal coercion and preservation thresholds are no longer fixed, but instead respond to the evolving innovation frontier and internal liquidity conditions. The static cutoffs derived above thus map into time-varying reallocation rules, with the coercion margin initially wide—preserving a broad base of innovation—and later collapsing as short-term strategies lose relevance. This dynamic transition between coercion and liquidation plays a central role in shaping long-run growth paths and institutional responses to financial shocks.

\subsection{Feasibility and Corner Solutions}

An interior solution arises only when the liquidity shock is below a critical threshold. 
\begin{align}
\lambda_l \leq 1 &\quad \Rightarrow \quad \Delta L \leq \frac{X_1}{2}\left[(\theta_1^s - \theta_1^l) + (2\theta_1^l - \theta_1^s)(\frac{1+\Gamma}{\Gamma})^2\right], \label{eq:lambda_s_feasibility}
\end{align}
When the shock exceeds this threshold, the upper cutoff binds at $\lambda_l = 1$, and the conglomerate must rely solely on liquidation and coercion—no firms can be preserved on a long-term strategy. For a full derivation see \ref{app:feasibility}.\\

As long as the liquidity shock is positive, then there will not be a degenerate solution where \(\lambda_s = 0\). If the liquidity shock is large, the conglomerate cannot afford to preserve any firm on a long-term strategy. The upper cutoff binds at \( \lambda_l = 1 \), and the conglomerate solves for the minimum level of liquidation required to close the funding gap:
\begin{equation}
\Delta L = \frac{1}{2} X_1 (\theta_1^s - \theta_1^l) + \frac{1}{2} X_1 \theta_1^l \lambda_s^2 \label{eq:corner3_constraint}
\end{equation}

Solving yields:
\begin{equation}
\lambda_s = \sqrt{ \frac{2}{X_1 \theta_1^l} \left( \Delta L - \frac{1}{2} X_1 (\theta_1^s - \theta_1^l) \right) } \label{eq:corner3_lambda_k}
\end{equation}

This is the extreme end of reallocation policy: the conglomerate liquidates the weakest firms and coerces the rest, suspending all long-term investment. Such a strategy may arise during deep financial crises or when short-term liquidity needs outstrip the long-run gains from innovation. Importantly, coercion remains the preferred option wherever possible, even in this extreme case.


\section{Dynamic Growth Model}
\label{sec:growth}

\subsection{From Static Payoffs to Dynamic Growth}
\label{subsec:static_to_dynamic}

The previous section derived optimal cutoff rules for reallocating capital across firms in response to liquidity shocks, based on static productivity payoffs. I now embed those rules into a dynamic growth environment, where firm-level innovation strategies influence the long-run trajectory of aggregate productivity. This formulation departs from canonical endogenous growth models by embedding reallocation cutoffs that respond endogenously to evolving innovation payoffs. Unlike frameworks with exogenous frictions or static misallocation, this setup allows the optimal policy margins to shift dynamically with development.

The final goods sector is represented by a representative conglomerate that combines the outputs of all intermediate firms using a linear production aggregator:

\begin{equation}
    y_t =  \int_0^1 A_t(i) x_t(i) \, di
\end{equation}
where \( A_t(i) \) is the productivity of firm \( i \) and \( x_t(i) \) is its labor input. Labor supply is normalized.\footnote{A linear aggregator simplifies the analysis by allowing aggregate output to scale directly with the weighted average of firm-level productivity. While CES aggregators with imperfect substitutability are standard in models with monopolistic competition or price dispersion (e.g., \citet{acemoglu2018micro}), my focus is on internal capital allocation and productivity dynamics, not on price-setting or demand-side reallocation. Moreover, allowing for imperfect substitutability would introduce additional curvature and demand shifters without changing the core mechanism of reallocation under financial constraints.}

Intra-period innovation choices determine the evolution of \( A_t(i) \). Each firm begins the period with the productivity level \( \lambda(i) \), drawn from a uniform distribution. At the beginning of the period, the firms all choose a strategy \( h \in \{s, l\} \): short-term (imitation) or long-term (frontier innovation). Using the setup from the previous section, each stage of the research process is given by:
\begin{align*}
    X_1^h(i) &= \lambda \theta_1^h X_1, \\
    X_2^h(i) &= \lambda \theta_2^h X_2,
\end{align*}

\begin{center}
    \begin{minipage}{0.45\textwidth}
    \textbf{Stage 1 Innovation Payoffs}
    \begin{align*}
    X_1 &= \beta \\
    \theta_1^s &= \overline{A}_{t-1} \\
    \theta_1^l &= A_{t-1}
    \end{align*}
    \end{minipage}
    \hfill
    \begin{minipage}{0.45\textwidth}
    \textbf{Stage 2 Innovation Payoffs}
    \begin{align*}
    X_2 &= 1 \\
    \theta_2^s &= \beta^s \overline{A}_{t-1} \\
    \theta_2^l &= \beta^l A_{t-1}
    \end{align*}
    \end{minipage}
\end{center}

This means the first round payoffs which determine liquidity value are given by: 
\begin{align}
    X_1^s(i) &= \lambda \beta \overline{A}_{t-1}, \\
    X_1^l(i) &= \lambda \beta A_{t-1}, 
\end{align}

The second round payoffs, which will define the final productivity for the firm, are given by:
\begin{align}
    X_2^s(i) &= \lambda \beta^s \overline{A}_{t-1}, \\
    X_2^l(i) &= \lambda \beta^l A_{t-1}.
\end{align}

I also assume that liquidity needs and hence the liquidity shocks scales with the size of the frontier and aggregate technology, so that:
\begin{align}
    \Delta L_t &= \eta \overline{A}_{t-1}
\end{align}

\( X_1 \) and \( X_2 \) are common scaling constants. The short-term strategy depends on \( \overline{A}_{t-1} \) (the exogenous frontier), while the long-term strategy builds from the conglomerate's own aggregate productivity \( A_{t-1} \). The liquidity returns (first-stage payoffs) differ across strategies by the coefficient \( \beta \) as well as the distance to the innovation frontier. 

\subsection{Aggregate Cutoff Policy}
\label{subsec:agg_cutoff}

The cutoff rules \( (\lambda_s, \lambda_l) \) are inherited from the conglomerate’s static problem but now evolve endogenously with the structure of innovation payoffs. Firms are still categorized into liquidation (\( \lambda < \lambda_s \)), coercion (\( \lambda_s \leq \lambda < \lambda_l \)), or long-termism (\( \lambda \geq \lambda_l \)), but the returns to each strategy now depend on firm productivity, the economy’s distance to the frontier, and the evolution of aggregate technological potential. As I show, this structure causes the liquidation and coercion cutoffs to converge endogenously as the economy approaches the frontier.

The conglomerate’s optimal cutoffs continue to satisfy the static solutions derived in Section~\ref{sec:model}:
Beginning with the ratio of the cutoff conditions: 
\begin{align}
    \Gamma &= \frac{(\theta_2^l - \theta_2^s)\theta_1^l}{(\theta_1^s - \theta_1^l)\theta_2^l} \nonumber\\
     &= \frac{(\beta^l A_{t-1} - \beta^s \overline{A}_{t-1})}{(\overline{A}_{t-1} - A_{t-1})} 
\end{align} 

Define the distance to the frontier as:
\begin{equation}
    a_{t} = \frac{A_{t}}{\overline{A}_{t}}\leq1
\end{equation}

Assuming that the distance to the frontier begins above a certain threshold $a_{t}\geq\beta$ (to ensure that long termism is the efficient choice outside of liquidity shocks), I have thus that 
\begin{align}
    \Gamma &= \frac{\beta^l a_{t-1}-\beta^s}{1-a_{t-1}} 
\end{align} 

As in Section~\ref{sec:model}, this allows me to define the relative margins of coercion and liquidation:
\begin{equation*}
    \lambda_l = \frac{1 + \Gamma}{\Gamma} \lambda_s
\end{equation*}

 This dynamic cutoff convergence reflects the fact that coercion yields diminishing marginal liquidity as imitation becomes less productive. Eventually, the conglomerate reallocates capital strictly between liquidation and long-term preservation, bypassing intermediate strategies altogether. 
 
 This expression implies as development progresses and \( a \to 1 \), then \( \Gamma \to \infty \), and therefore the cutoffs converge and the coercion margin shrinks to zero:
\begin{equation}
\lambda_l \to \lambda_s,
\end{equation}

This implies that the coercion margin disappears. This convergence reflects an improvement in allocation efficiency, as capital is increasingly concentrated in high-potential firms when short-termist strategies lose their liquidity advantage. This dynamic transition—from three margins to two—offers a tractable alternative to models with exogenously fixed firm types or permanent distortions, such as those in \citet{buera2011financial}, by making firm allocation a dynamic function of the innovation environment and conglomerate liquidity needs.

While cutoff convergence implies the conglomerate eventually abandons coercion, it does not mean all firms are preserved if liquidity shocks scale with the economy. Looking at the liquidation margin:

\begin{align}
    \lambda_s = \sqrt{ \frac{2 \eta}
    {\beta  \left[(1-a_{t-1})\left(\frac{1 + \Gamma}{\Gamma}\right)^2 + (2a_{t-1} - 1)\right]}}
\end{align}

The parameter \( \eta \) governs the scale of the liquidity shock relative to the aggregate productivity frontier \( \overline{A}_{t-1} \), such that \( \Delta L_t = \eta \overline{A}_{t-1} \). The full derivation is given in \ref{appsub:liquidation_margin}.\\

Economically, \( \eta \) captures the intensity of systemic funding pressures faced by the conglomerate during development—for example, the extent of debt refinancing needs, rollover risk, or external funding shortfalls that rise with firm size or macroeconomic complexity. In highly financialized or credit-constrained economies, these liquidity needs often scale with overall output or capital stock. For instance, models of financial amplification such as Bernanke, Gertler, and Gilchrist (1999) and Buera, Kaboski, and Shin (2011) emphasize how aggregate shocks propagate more forcefully when firm leverage or external dependence rises with development. Empirically, Kalemli-Özcan et al.\ (2015) and Greenwald, Lettau, and Ludvigson (2019) document that large firms and conglomerates face liquidity constraints that are increasing in scale. Modeling \( \Delta L_t \) as proportional to \( \overline{A}_{t-1} \) via \( \eta \) thus reflects the realistic assumption that the funding burden a conglomerate faces grows with its size and sophistication. 

Note that as development progresses and \( a \to 1 \) and \( \Gamma \to \infty \), then 
\begin{equation}
    \lambda_s \to \sqrt{\frac{2\eta}{\beta}} \label{eq:lambda_s_limit}
\end{equation}

Importantly, the model shows that even as short-termism vanishes (\( \lambda_l \to \lambda_s \)), a strictly positive liquidation margin persists if \( \eta > 0 \), since the group must continue to raise internal liquidity proportional to aggregate scale.

When the economy is far from the frontier (\( a \ll 1 \)), short-term imitation is highly effective. In this case, the marginal liquidity gain from coercion is large, and the conglomerate can satisfy its funding constraint by reallocating intermediate firms. The cutoff ratio \( \Gamma \to 0 \), and the denominator diverges, causing \( \lambda_s \to 0 \). In other words, almost no firms are liquidated when imitation remains productive. As the economy develops and \( a \to 1 \), the value of imitation collapses. The marginal liquidity gain from coercion vanishes, and the conglomerate increasingly relies on liquidation to satisfy its funding constraint. In this case, \( \Gamma \to \infty \), and the cutoff simplifies to \eqref{eq:lambda_s_limit}

Thus, while the coercion region \( [\lambda_s, \lambda_l) \) vanishes in the limit, the liquidation margin persists. The conglomerate drops a strictly positive share of low-\( \lambda \) firms even near the frontier, reflecting the growing inefficiency of preserving marginal firms once short-term payoffs deteriorate. This shift underpins the model’s core prediction: optimal reallocation evolves from coercion toward a binary regime of liquidation versus long-term investment as the economy approaches the frontier. At the frontier, if liquidity shocks fall in relative size, the conglomerate can afford to preserve more and more firms. Unlike models that assume exogenous innovation types or static firm heterogeneity, this framework shows how firm strategy sets and capital allocations co-evolve with development.

\subsection{Strategic Allocation by Development Level}
\label{subsec:cutoff_shock_panels}

\begin{figure}[H]
    \centering
    \includegraphics[width=\textwidth]{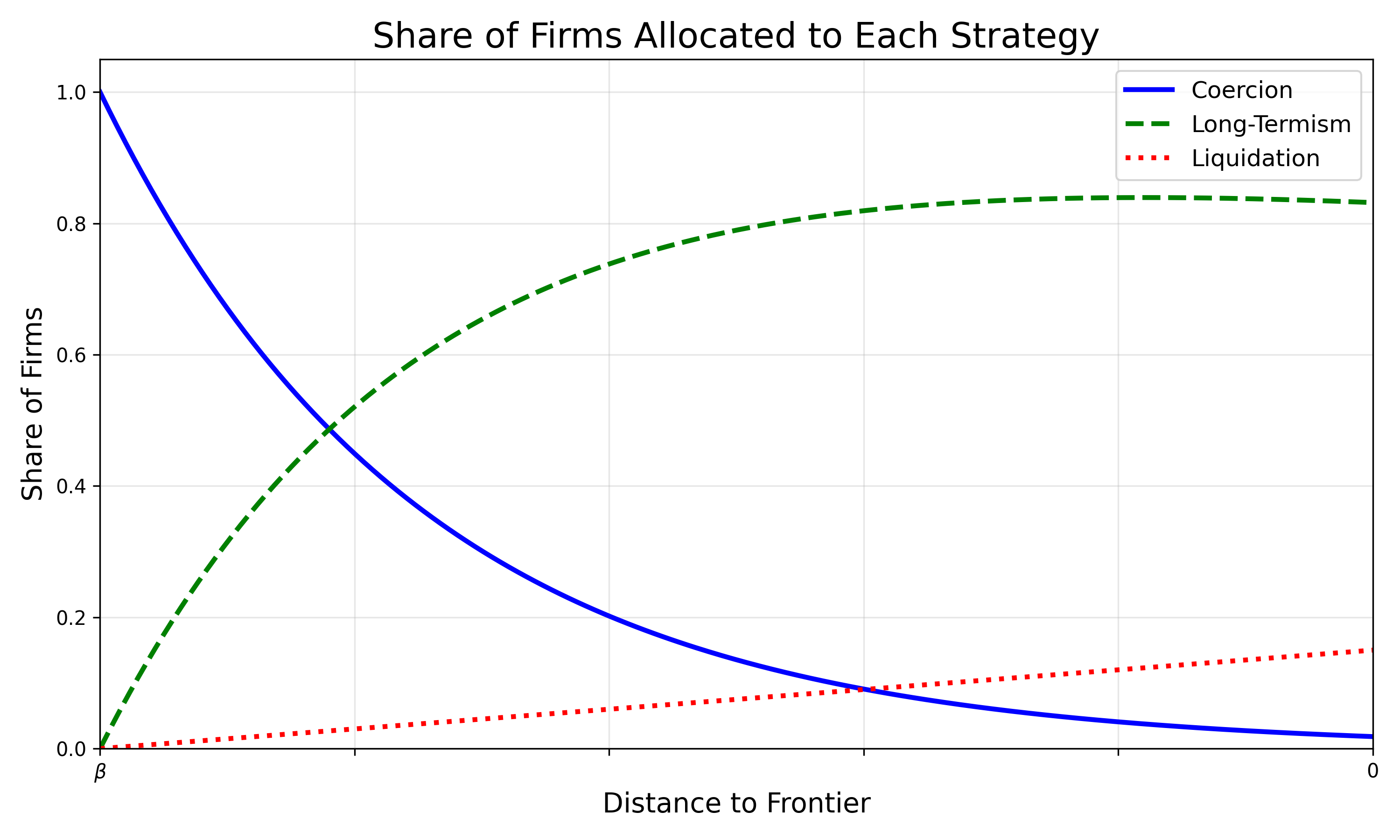}
    \caption{Strategic Allocation by Development Level. Each panel shows the share of firms allocated to liquidation (red), coercion (blue), and long-termism (green) as the economy approaches the frontier. Imitation loses effectiveness at high development levels, prompting reallocation away from coercion. Larger shocks sustain reliance on liquidation.}
    \label{fig:cutoff_shock_panels}
\end{figure}

As the economy advances and firms approach the global technology frontier, the conglomerate’s internal capital allocation adapts endogenously. The erosion of imitation returns shifts the liquidity-efficiency trade-off that governs firm reallocation. This causes the interior cutoffs \( (\lambda_s, \lambda_l) \) to converge over time, gradually phasing out the intermediate coercion margin while retaining a positive liquidation margin. Figure~\ref{fig:cutoff_shock_panels} illustrates the conglomerate’s evolving reallocation strategy across three levels of financial stress. Each panel shows the share of firms assigned to liquidation, coercion, or long-term R\&D, as a function of distance to the technology frontier.

\subsection{Strategic Transitions and Dynamic Misallocation}
\label{subsec:strategic_transition}

The convergence of cutoffs \( \lambda_s \) and \( \lambda_l \) implies that coercion to short-termism, while useful during early development, must fade as the economy advances. As imitation loses effectiveness, the conglomerate's ability to use intermediate reallocation strategies erodes. The conglomerate must transition toward a binary allocation between liquidation and long-term preservation. The coercion region \( [\lambda_s, \lambda_l) \) shrinks to zero, leaving a strictly positive liquidation share even near the frontier. This endogenous shift reflects the conglomerate’s adaptation to declining short-term innovation value.

Dynamic misallocation can arise not merely from financial shocks, but from institutional inertia—the failure to adjust strategy margins in response to a changing innovation landscape. This endogenous strategic shift mirrors historical reallocation failures observed in East Asian bank-based economies, where conglomerates continued coercive reallocation long after imitation had lost its developmental edge (See \citet{Branstetter2002HasJI,hoshi2001corporate}). In the next section, I characterize the long-run growth losses associated with such rigidity.


\section{Growth and Dynamic Misallocation}
\label{sec:policy}

In this section I examine how aggregate growth evolves under the dynamic reallocation structure developed in Section~\ref{sec:growth}, and how persistent misallocation arises when the conglomerate maintains suboptimal strategy margins over time. The erosion of short-term innovation returns causes cutoffs to evolve, raising both the share of capital maintained for long-term innovation and the share of firms sacrificed to provide liquidity for those long-termist firms after a liquidity shock. If reallocation margins do not adjust accordingly, growth suffers not because of shocks alone, but because of institutional rigidity.

\subsection{Aggregate Growth Dynamics} \label{subsec:agg_growth}

Given the production structure in Section~\ref{sec:growth}, aggregate productivity growth is the average across surviving firms. Let cutoffs \( \lambda_s \) and \( \lambda_l \) define the regions for liquidation, coercion to short-termism, and preservation for long-term investment. 

Aggregate growth in period \( t \) is:
\begin{align}
   1+g_t &= \int_0^1 \left( \frac{A_t(i)}{A_{t-1}(i)}  \right) di \nonumber \\
         &= \int_{\lambda_s}^{\lambda_l} \left( \frac{\lambda \theta^s_2 X_2}{A_{t-1}}  \right)d\lambda + \int_{\lambda_l}^1 \left( \frac{\lambda \theta^l_2 X_2}{A_{t-1}}  \right)d\lambda \nonumber\\
         &= \int_{\lambda_s}^{\lambda_l} \left( \frac{\lambda \beta^s \bar{A}_{t-1}}{A_{t-1}}  \right)d\lambda + \int_{\lambda_l}^1 \left( \frac{\lambda \beta^s A_{t-1}}{A_{t-1}}  \right)d\lambda \nonumber\\
         &= \frac{(\lambda_l^2-\lambda_s^2)}{2}\frac{\beta^s}{a_t}+ \frac{(1-\lambda_l^2)}{2}\beta^l
\end{align}

The aggregate distance to the frontier evolves as: 
\begin{align}
    a_t &= \int_0^1 \left( \frac{A_t(i)}{\bar{A}_{t-1}(i)}  \right) di \nonumber \\
    &= \int_{\lambda_s}^{\lambda_l} \left( \frac{\lambda \theta^s_2 X_2}{\bar{A}_{t-1}}  \right)d\lambda + \int_{\lambda_l}^1 \left( \frac{\lambda \theta^l_2 X_2}{A_{t-1}}  \right)d\lambda \nonumber\\
    &= \int_{\lambda_s}^{\lambda_l} \left( \frac{\lambda \beta^s \bar{A}_{t-1}}{\bar{A}_{t-1}}  \right)d\lambda + \int_{\lambda_l}^1 \left( \frac{\lambda \beta^s A_{t-1}}{\bar{A}_{t-1}}  \right)d\lambda \nonumber\\
    &= \frac{(\lambda_l^2-\lambda_s^2)}{2}\beta^s+ \frac{(1-\lambda_l^2)}{2}\beta^la_{t-1}
\end{align}

Growth from short termism declines as \( a_t \to 1 \). The conglomerate’s strategy must adapt over time to favor long-termist firms as the growth differential narrows.

Remember that the interior solution is given by: 
\begin{equation*}
    \lambda_l = \frac{1 + \Gamma}{\Gamma} \lambda_s
\end{equation*}

where
\begin{align*}
    \Gamma &= \frac{\beta^l a_{t-1}-\beta^s}{1-a_{t-1}} 
\end{align*} 

and 
\begin{align*}
    \lambda_s 
    &= \sqrt{ \frac{2 \eta}
    {\beta  \left[(1-a_{t-1})\left(\frac{1 + \Gamma}{\Gamma}\right)^2 + (2a_{t-1} - 1)\right]}}
\end{align*}

\subsection{Counterfactual Reallocation Regimes: Coercion-Only vs. Liquidation-Only}
\label{subsec:counterfactual_regimes}

To assess how internal reallocation strategies shape long-run growth, I consider two counterfactual regimes: one in which the conglomerate relies exclusively on coercion (coercion-only), and one in which it relies solely on liquidation (liquidation-only). These regimes are not optimal policies, but stylized extremes that illustrate how the relative benefits of each margin evolve with development. Comparing these paths to the optimal allocation helps identify when coercion is useful, when liquidation is preferable, and where misallocation arises.

\subsection{The Coercion Trap}
\label{subsec:coercion_trap}

In the coercion-only regime, the conglomerate never liquidates firms. Liquidity constraints are met entirely by coercing firms into short-termist strategies, setting \( \lambda_s = 0 \). 

The cutoff \( \lambda_l^c \) is determined by the liquidity constraint:
\begin{align}
\Delta L &= \frac{1}{2} (\theta_1^s - \theta_1^l) X_1 \lambda_l^2.
\end{align}

Solving in terms of the general payoff structure of Section \ref{sec:model} and substituting the particular payoffs from Section \ref{sec:growth} yields:
\begin{align}
\lambda_l^c &= \sqrt{ \frac{2\Delta L}{X_1 (\theta_1^s - \theta_1^l)} } = \sqrt{ \frac{2\eta}{\beta (1 - a_{t-1})} }.
\end{align}

Since firms are indexed over \([0,1]\), the cutoff cannot exceed one. Define the capped threshold:
\begin{align}
\tilde{\lambda}_l^c = \min\left\{1, \lambda_l^c \right\}.
\end{align}

Under this strategy, aggregate growth is:
\begin{align}
1 + g_c(a) &= \frac{1}{2a} \left( \tilde{\lambda}_l^{c\,2} \beta^s + (1 - \tilde{\lambda}_l^{c\,2}) \beta^l a \right). \label{eq:gc_cap}
\end{align}

In contrast, the conglomerate’s optimal allocation yields:
\begin{align}
1 + g^*(a) &= \frac{1}{2a} \left( (\lambda_l^{*2} - \lambda_s^{*2}) \beta^s + (1 - \lambda_l^{*2}) \beta^l a \right). \label{eq:gs_opt}
\end{align}

The growth loss is:
\begin{align}
\text{Loss}_c(a) &= g^*(a) - g_c(a). \label{eq:loss_def}
\end{align}

\paragraph{Sign of the Loss}

Letting \( \lambda_l^c \leq 1 \), note that:
\begin{align}
\lambda_s^{*2} > 0, \quad \lambda_l^{*2} < \tilde{\lambda}_l^{c\,2}, \quad \text{and} \quad \frac{\beta^s}{a} > \beta^l,
\end{align}

Both components of the growth gap are strictly positive. Hence for the entire range of development levels up to the technological frontier, the optimal strategy is strictly preferred:
\begin{align}
\text{Loss}_c(a) > 0 \quad \text{for all} \quad a \in (\beta, 1).
\end{align}

\paragraph{Limit Behavior} At the limits of the range of development, the growth loss converges to zero or a strictly positive finite value. The conservative strategy is as good as the optimal strategy only when the economy is as far as possible from the frontier. At the other extreme, the conservative strategy is strictly worse than the optimal strategy. 

As \( a \to \beta \) (the farthest distance to the frontier):
\begin{align}
\lambda_s^* &\to 0, \quad \Gamma \to 0, \quad \lambda_l^* \to \lambda_l^c, \nonumber\\
\Rightarrow \quad \text{Loss}_c(a) &\to 0.
\end{align}

As \( a \to 1 \) (the technological frontier), the uncapped \( \lambda_l^c \to \infty \), but is bounded at \( \tilde{\lambda}_l^c = 1 \). Therefore:
\begin{align}
1 + g_c(a) &\to \frac{\beta^s}{2} ,\nonumber \\
1 + g^*(a) &\to \frac{\beta^l}{2} ,\nonumber \\
\Rightarrow \quad \text{Loss}_c(a) &\to \frac{(\beta^l - \beta^s)}{2}.
\end{align}

The loss remains finite and strictly positive in the limit. Full derivations are included in \ref{appsub:coercion_trap_limit}.

This regime performs well early in development, when imitation yields strong returns. But near the frontier, coercion becomes inefficient, and growth suffers relative to the optimal policy.

\subsection{The Liquidation Fallacy}
\label{subsec:liquidation_fallacy}

In contrast, the liquidation-only regime forbids coercion, setting \( \lambda_l = \lambda_s \). The conglomerate drops firms to meet its liquidity needs and allocates the remainder to long-term strategies. 

\begin{align}
\Delta L &= \frac{1}{2} (\theta_1^s - \theta_1^l) X_1 \lambda_s^2.
\end{align}

Solving yields:
\begin{align}
\lambda_s^\ell &= \sqrt{ \frac{2\Delta L}{X_1 (\theta_1^s - \theta_1^l)} } = \sqrt{ \frac{2\eta}{\beta (1 - a_{t-1})} }, \\
\lambda_l^\ell &= \lambda_s^\ell.
\end{align}

Imposing the upper bound, define:
\begin{align}
\tilde{\lambda}_s^\ell = \min\{1, \lambda_s^\ell\}.
\end{align}

Under this liquidation-only strategy, all surviving firms pursue long-term innovation. Aggregate growth is:
\begin{align}
1 + g_\ell(a) &= \frac{1}{2a} \left( (1 - \tilde{\lambda}_s^{\ell\,2}) \beta^l a \right) = \frac{1}{2} (1 - \tilde{\lambda}_s^{\ell\,2}) \beta^l.
\end{align}

In contrast, the conglomerate’s optimal allocation allows coercion when imitation remains effective:
\begin{align}
1 + g^*(a) &= \frac{1}{2a} \left( (\lambda_l^{*2} - \lambda_s^{*2}) \beta^s + (1 - \lambda_l^{*2}) \beta^l a \right).
\end{align}

Define the growth loss from the liquidation fallacy as:
\begin{align}
\text{Loss}_\ell(a) &= g^*(a) - g_\ell(a).
\end{align}

\paragraph{Sign of the Loss}

When \( \lambda_l^* > \lambda_s^* \) and \( \lambda_s^{*2} < \tilde{\lambda}_s^{\ell\,2} \), the conglomerate’s strategy retains more productive coercible firms than the liquidation-only strategy. Since \( \frac{\beta^s}{a} > \beta^l \) in early development, I have:
\begin{align}
\text{Loss}_\ell(a) > 0 \quad \text{for all} \quad a \in (\beta, 1),
\end{align}

Just like with the coercion trap, the liquidation fallacy is strictly positive for all \( a \) strictly within  the development range. 

\paragraph{Limit Behavior} The growth loss converges to zero or a strictly positive finite value at the limits of the development range. The overly aggressive strategy, however, converges to the optimal strategy at the frontier.

As \( a \to \beta \), I have:
\begin{align}
\text{Loss}_\ell(a) \to \frac{1}{2\beta} \left[ \beta^l \beta + \frac{2\eta}{1 - \beta} \left( \frac{\beta^s}{\beta} - \beta^l \right) \right],
\end{align}

which is increasing in \( \eta \) and strictly positive as long as \( \beta^s > \beta \beta^l \). This captures the inefficiency of premature liquidation during early development.

As \( a \to 1 \), I have:
\begin{align}
\text{Loss}_\ell(a) \to 0.
\end{align}

Full derivations are included in \ref{appsub:liquidation_fallacy_limit}.

This regime is more effective near the frontier, when imitation has little marginal value. But in early development, it discards too many productive firms, leading to excessive misallocation and lower growth.

\subsection{Comparative Growth Under Reallocation Regimes}
\label{subsec:growth_graph}

\begin{figure}[H]
    \centering
    \includegraphics[width=0.85\textwidth]{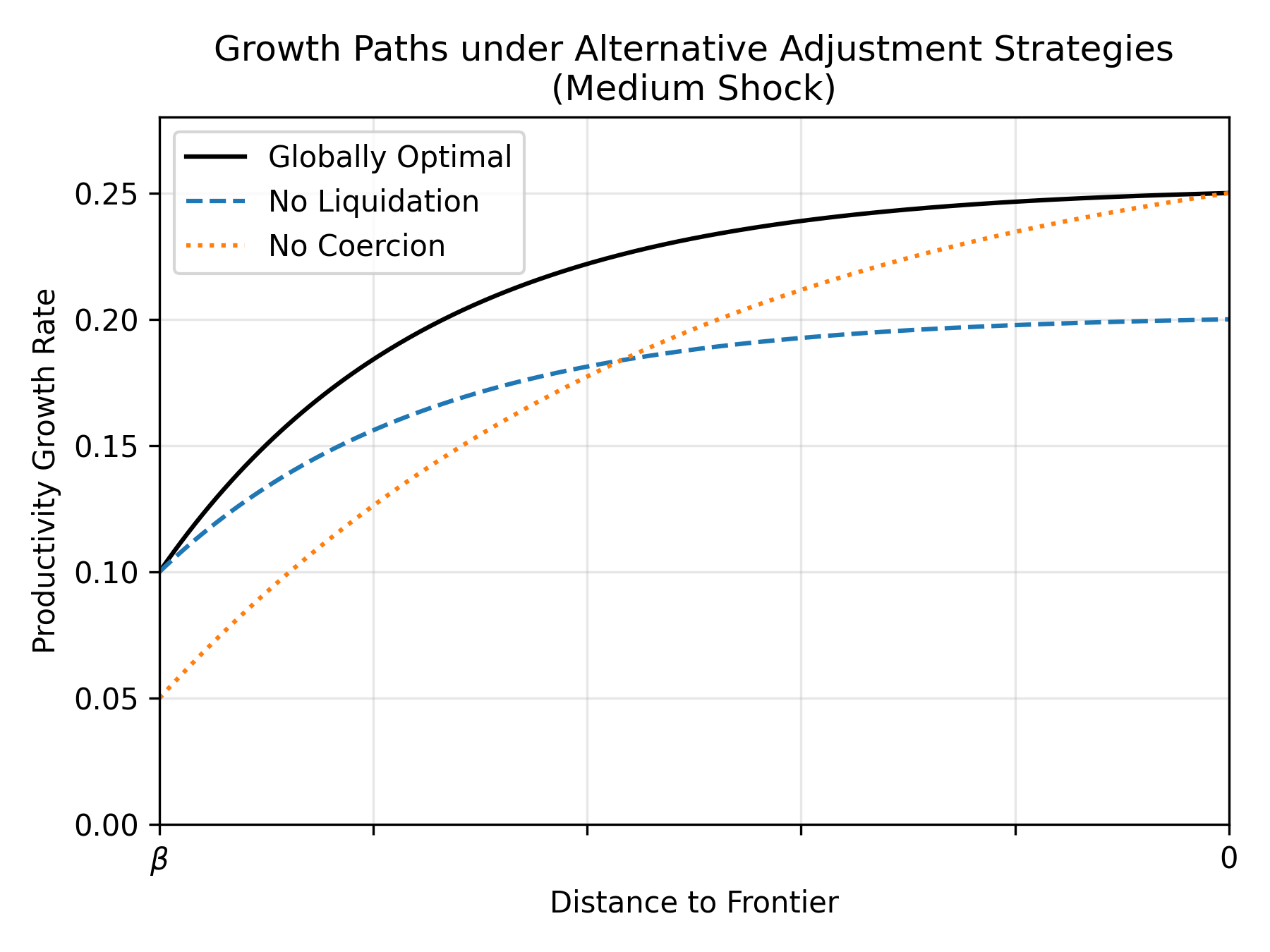}
    \caption{Productivity Growth Under Alternative Adjustment Strategies. The fully flexible conglomerate (black) transitions from coercion to long-termism. Coercion-only (blue) stagnates near the frontier. Liquidation-only (red) underperforms early in development.}
    \label{fig:growth_paths}
\end{figure}

Figure~\ref{fig:growth_paths} plots all three strategies. The optimal policy transitions smoothly: coercion dominates early, liquidation rises with development, and both margins shrink near the frontier. The counterfactual regimes deviate in opposite directions. The coercion-only regime underperforms late, while the liquidation-only regime underperforms early. This pattern underscores that the value of each strategy is development-dependent.

These counterfactuals illustrate the costs of rigid reallocation regimes. The coercion-only strategy becomes a \textit{coercion trap} when sustained too long, tying capital to firms that can no longer deliver growth. The liquidation-only strategy embodies a \textit{liquidation fallacy}, discarding too much capacity in pursuit of early selection. Neither policy is always inferior—but both are context-sensitive. Effective capital allocation depends not only on which margin is used, but on when and how flexibly it is applied.

\subsection{Policy Implications and Broader Lessons}
\label{subsec:policy}

The model highlights how persistent growth slowdowns can stem not from liquidity shocks per se, but from institutional rigidity—a failure to adapt internal capital reallocation as the economy develops. Fixed strategy margins, even if ex ante efficient, become distortive when held constant across stages of development, especially as the relative return to long-run innovation rises.

To illustrate this point, I compare optimal reallocation with two stylized extremes: a coercion trap, where all firms are preserved through short-run strategies that eventually lose effectiveness; and a liquidation fallacy, where firms are prematurely discarded to conserve liquidity despite long-run potential. Both regimes lead to persistent misallocation and slower growth—not due to classic borrowing frictions, but because reallocation strategies fail to evolve with innovation payoffs.

These dynamics are especially salient in bank-based and conglomerate-led economies. South Korea’s post-crisis restructuring of its chaebol system involved state-led liquidation and recapitalization to reallocate capital toward high-potential firms \citep{chang2001interpretkorean, joh2003corporate}. In contrast, Japan’s lost decade reflects a coercion trap: keiretsu networks and weak banks sustained unproductive firms, contributing to stagnation \citep{hoshi2001corporate, peek2005unnatural}. In both cases, institutional flexibility—or the lack thereof—shaped internal capital market effectiveness.

These insights also resonate beyond East Asia. In emerging markets, political or relational lending systems may preserve unproductive firms for non-economic reasons. But aggressive liquidation regimes—e.g., those inspired by Washington Consensus reforms—may err in the other direction, discarding firms before their innovation potential matures \citep{rodrik2006goodbye, stiglitz2002globalization}. In both extremes, growth suffers when capital reallocation fails to keep pace with changing innovation dynamics.

In sum, the model suggests that reallocation flexibility—not just scale or discipline—is central to innovation-led growth. Institutions that adapt strategy margins to the evolving innovation landscape can sustain growth; those that do not risk locking into inefficient paths.


\section{Conclusion}

I develop a dynamic model of internal capital allocation in conglomerates facing conglomerate-level liquidity shocks. I derive cutoff rules for preservation, coercion, and liquidation based on affiliate firm productivity heterogeneity. I embed these rules in an endogenous growth framework and show that as the economy advances toward the global technology frontier that optimal reallocation strategies evolve. Coercion becomes less effective and long-term innovation plays a growing role in sustaining growt, both because of increased payoffs of long-term innovation but also because of the declining liquidity value of imitation. I examine two stylized forms of misallocation. In a a coercion trap, all firms are preserved and so more are pushed into short-term strategies that become less effective as the economy develops. In a liquidation fallacy, no firms are coerced and so too many firms are discarded for liquidity. The former is increasingly suboptimal as the economy approaches the frontier, while the latter is suboptimal in early development. 

Future work could extend the model to include overlapping generations of firms or persistent financial frictions, enabling more realistic survival dynamics and policy counterfactuals. Embedding the framework in a normative environment—with explicit welfare or volatility objectives—would clarify optimal policy design. Bailouts, restructuring programs, and innovation subsidies interact with internal strategy rules in nontrivial ways. Effective intervention thus requires aligning external support with evolving internal strategy margins. Cross-country extensions could examine how institutional features, such as conglomerate-bank linkages or regulatory regimes, shape reallocation flexibility across development paths. Policies that preserve firms may support growth when short-term survival is essential but could become harmful when long-run innovation drives the economy. 

The model provides a structured way to interpret and measure how conglomerates reallocate capital under financial stress—and how the payoff to different strategies shifts with development. By observing how firms are preserved, coerced, or liquidated across time and productivity distributions, one can recover the evolving shape of the underlying payoff functions. This could be done through tracking patterns in R\&D persistence, follow-on investment, and growth outcomes. Observed cutoff behavior offers a natural basis for estimating how sensitive strategy selection is to firm productivity or distance from the technological frontier. Over time, shifts in these empirical margins can be interpreted as adaptations to changing innovation environments, not necessarily as signs of inefficiency. In this way, the model could be the beginning of an empirical agenda centered on identifying how internal capital markets process information and respond to evolving tradeoffs—yielding insights into the dynamics of growth under conglomerates.

Ultimately, I show that innovation-led growth depends not just on external funding or internal control, but on the ability to adapt reallocation strategies as the economy evolves. In a world of fragile innovation, frequent financial shocks, and increasingly large conglomerates, understanding how capital flows across firms and over time is essential. This paper offers one step toward that goal.

\newpage


\appendix
\section{Derivations and Cutoff Conditions}
\renewcommand{\theequation}{A.\arabic{equation}}
\setcounter{equation}{0}

\subsection{Derivation of Interior Cutoff Condition} \label{app:interior_cutoff}

The conglomerate’s objective is:
\begin{equation}
\max_{\lambda_s, \lambda_l} \int_{\lambda_s}^{\lambda_l} X_2^s(\lambda)\, d\lambda + \int_{\lambda_l}^1 X_2^l(\lambda)\, d\lambda
\end{equation}

subject to the liquidity constraint:
\begin{equation}
\Delta L = \int_0^{\lambda_s} X_1^l(\lambda)\, d\lambda + \int_{\lambda_s}^{\lambda_l} \left[X_1^s(\lambda) - X_1^l(\lambda)\right]\, d\lambda
\end{equation}

and given the functional forms:
\begin{align}
X_1^h(\lambda) = \lambda \theta_1^h X_1, \quad X_2^h(\lambda) = \lambda \theta_2^h X_2,
\end{align}

The integrals within the objective function can thus be rewritten as:
\begin{align}
\int_{\lambda_s}^{\lambda_l} \lambda \theta_2^s X_2\, d\lambda &= \theta_2^s X_2 \cdot \frac{1}{2} (\lambda_l^2 - \lambda_s^2) \\
\int_{\lambda_l}^{1} \lambda \theta_2^l X_2\, d\lambda &= \theta_2^l X_2 \cdot \frac{1}{2}(1 - \lambda_l^2)
\end{align}

The integrals within the liquidity constraint can thus be rewritten as:
\begin{align}
\int_0^{\lambda_s} \lambda \theta_1^l X_1\, d\lambda &= \theta_1^l X_1 \cdot \frac{1}{2} \lambda_s^2 \\
\int_{\lambda_s}^{\lambda_l} \lambda (\theta_1^s - \theta_1^l) X_1\, d\lambda &= (\theta_1^s - \theta_1^l) X_1 \cdot \frac{1}{2} (\lambda_l^2 - \lambda_s^2)
\end{align}

So total liquidity is given by 
\begin{equation}
\Delta L = \frac{1}{2} X_1 \left[ (\theta_1^s - \theta_1^l) \lambda_l^2 + (2\theta_1^l - \theta_1^s) \lambda_s^2 \right]
\end{equation}

To find the optimal interior cutoff condition, substitute the constraint and use a Lagrangian:
\begin{align}
\mathcal{L} =& \frac{1}{2} \theta_2^s X_2 (\lambda_l^2 - \lambda_s^2) + \frac{1}{2} \theta_2^l X_2 (1 - \lambda_l^2) \nonumber\\
&+ \mu \left[ \Delta L - \frac{1}{2} X_1 \left((\theta_1^s - \theta_1^l) \lambda_l^2 + (2\theta_1^l - \theta_1^s) \lambda_s^2 \right) \right]
\end{align}

FOCs:
\begin{align}
\frac{\partial \mathcal{L}}{\partial \lambda_l} = \theta_2^s X_2 \lambda_l - \theta_2^l X_2 \lambda_l - \mu X_1 (\theta_1^s - \theta_1^l) \lambda_l = 0\\
\frac{\partial \mathcal{L}}{\partial \lambda_s} = -\theta_2^s X_2 \lambda_s - \mu X_1 (2\theta_1^l - \theta_1^s) \lambda_s = 0
\end{align}

Divide the two conditions to eliminate \( \mu \):
\begin{align}
\frac{\theta_2^l - \theta_2^s}{\theta_2^s} &= \frac{(\theta_1^s - \theta_1^l)}{(2\theta_1^l - \theta_1^s)} \cdot \frac{\lambda_s}{\lambda_l} \label{app:lambda_ratio}
\end{align}
    
Define:
\begin{align}
\Gamma \equiv \frac{(\theta_2^l - \theta_2^s)\theta_1^l}{(\theta_1^s - \theta_1^l)\theta_2^l}
\end{align}

First, rearrange equation \ref{app:lambda_ratio} to solve for \( \frac{\lambda_s}{\lambda_l} \):
\begin{align}
\frac{\lambda_s}{\lambda_l} = \frac{\theta_2^l - \theta_2^s}{\theta_2^s} \cdot \frac{(2\theta_1^l - \theta_1^s)}{(\theta_1^s - \theta_1^l)}
\end{align}

Now express the right-hand side in terms of \( \Gamma \). Multiply numerator and denominator by \( \theta_1^l/\theta_2^l \):
\begin{align}
\frac{\lambda_s}{\lambda_l} = \left( \frac{(\theta_2^l - \theta_2^s)\theta_1^l}{(\theta_1^s - \theta_1^l)\theta_2^l} \right) \cdot \left( \frac{2\theta_1^l - \theta_1^s}{\theta_1^l} \right)
= \Gamma \cdot \left( \frac{2\theta_1^l - \theta_1^s}{\theta_1^l} \right)
\end{align}

Now simplify:
\begin{align}
\frac{\lambda_s}{\lambda_l} = \Gamma \cdot \left( \frac{2\theta_1^l - \theta_1^s}{\theta_1^l} \right) = \Gamma \left(2 - \frac{\theta_1^s}{\theta_1^l} \right)
\end{align}

Let’s now isolate \( \lambda_l \):
\begin{align}
\frac{\lambda_l}{\lambda_s} = \frac{1}{\Gamma \left(2 - \frac{\theta_1^s}{\theta_1^l} \right)}
\end{align}

But from the definition of \( \Gamma \), I can solve for \( \frac{\theta_1^s}{\theta_1^l} \) in terms of \( \Gamma \). Start by multiplying both sides by denominator:
\begin{align}
\Gamma (\theta_1^s - \theta_1^l)\theta_2^l = (\theta_2^l - \theta_2^s)\theta_1^l
\Rightarrow \Gamma \theta_1^s \theta_2^l - \Gamma \theta_1^l \theta_2^l = (\theta_2^l - \theta_2^s)\theta_1^l
\end{align}

Rewriting:
\begin{align}
\Gamma \theta_1^s \theta_2^l = \Gamma \theta_1^l \theta_2^l + (\theta_2^l - \theta_2^s)\theta_1^l
\Rightarrow \theta_1^s = \theta_1^l \left(1 + \frac{(\theta_2^l - \theta_2^s)}{\Gamma \theta_2^l} \right)
\end{align}

So:
\begin{align}
\frac{\theta_1^s}{\theta_1^l} = 1 + \frac{1}{\Gamma} \cdot \left( \frac{\theta_2^l - \theta_2^s}{\theta_2^l} \right)
\end{align}

Hence:
\begin{align}
2 - \frac{\theta_1^s}{\theta_1^l} = 1 - \frac{1}{\Gamma} \cdot \left( \frac{\theta_2^l - \theta_2^s}{\theta_2^l} \right)
\end{align}

But since \( \Gamma = \frac{(\theta_2^l - \theta_2^s)\theta_1^l}{(\theta_1^s - \theta_1^l)\theta_2^l} \), this simplifies back down to:
\begin{align}
\frac{\lambda_l}{\lambda_s} = \frac{1 + \Gamma}{\Gamma}
\Rightarrow \lambda_l = \frac{1 + \Gamma}{\Gamma} \lambda_s
\end{align}

To simplify, redefine:
\begin{align}
\Gamma \equiv \frac{(\theta_2^l - \theta_2^s)\theta_1^l}{(\theta_1^s - \theta_1^l)\theta_2^l}
\quad \Rightarrow \quad
\lambda_l = \frac{1+\Gamma}{\Gamma} \lambda_s
\end{align}

\subsection{Closed-Form Expression for $\lambda_s$} \label{app:lambda_s}

I begin from the liquidity constraint:
\begin{equation}
\Delta L = \frac{1}{2} X_1 \left[ (\theta_1^s - \theta_1^l) \lambda_l^2 + (2\theta_1^l - \theta_1^s) \lambda_s^2 \right]
\end{equation}

Substitute the expression for \( \lambda_l \) in terms of \( \lambda_s \):
\begin{equation}
\lambda_l = \frac{1+\Gamma}{\Gamma} \lambda_s \quad \Rightarrow \quad \lambda_l^2 = \left( \frac{1+\Gamma}{\Gamma} \right)^2 \lambda_s^2
\end{equation}

Plug this into the liquidity constraint:
\begin{align}
\Delta L &= \frac{1}{2} X_1 \left[ (\theta_1^s - \theta_1^l) \left( \frac{1+\Gamma}{\Gamma} \right)^2 \lambda_s^2 + (2\theta_1^l - \theta_1^s) \lambda_s^2 \right] \nonumber \\
&= \frac{1}{2} X_1 \lambda_s^2 \left[ (\theta_1^s - \theta_1^l) \left( \frac{1+\Gamma}{\Gamma} \right)^2 + (2\theta_1^l - \theta_1^s) \right] 
\end{align}

Now isolate \( \lambda_s^2 \) by dividing both sides:
\begin{equation}
\lambda_s^2 = \frac{2\Delta L}{X_1 \left[ (\theta_1^s - \theta_1^l) \left( \frac{1+\Gamma}{\Gamma} \right)^2 + (2\theta_1^l - \theta_1^s) \right]}
\end{equation}

Finally, take the square root to solve for \( \lambda_s \):
\begin{equation}
\lambda_s = \sqrt{ \frac{2\Delta L}{X_1 \left[ (\theta_1^s - \theta_1^l) \left( \frac{1+\Gamma}{\Gamma} \right)^2 + (2\theta_1^l - \theta_1^s) \right]} }
\end{equation}

\subsection{Corner Case: Full Coercion, No Long-Term Investment} \label{app:cutoff large shock}

In this corner case, the conglomerate coerces all non-liquidated firms into short-term strategies and preserves none for long-term innovation. This implies \( \lambda_l = 1 \), so that no firm is left on the long-term strategy margin. The liquidity constraint becomes:
\begin{align}
\Delta L &= \int_0^{\lambda_s} X_1^l(\lambda) \, d\lambda + \int_{\lambda_s}^{1} (X_1^s(\lambda) - X_1^l(\lambda))\, d\lambda \nonumber \\
&= \frac{1}{2} X_1 \left[ \theta_1^l \lambda_s^2 + (\theta_1^s - \theta_1^l)(1 - \lambda_s^2) \right] \nonumber \\
&= \frac{1}{2} X_1 \left[ \theta_1^s - (\theta_1^s - \theta_1^l) \lambda_s^2 \right]
\end{align}

Solving for \( \lambda_s \), I obtain:
\begin{align}
\lambda_s^2 = \frac{2\Delta L - X_1 \theta_1^s}{X_1 (\theta_1^l - \theta_1^s)} \quad \Rightarrow \quad
\lambda_s = \sqrt{ \frac{2\Delta L - X_1 \theta_1^s}{X_1 (\theta_1^l - \theta_1^s)} }
\end{align}

This solution is valid only if:
\begin{equation}
\Delta L \geq \frac{1}{2} X_1 \theta_1^s
\end{equation}
which ensures that even after coercing all surviving firms, the conglomerate still needs to liquidate some share \( \lambda_s \) to meet the liquidity requirement.

Importantly, this corner case also requires that the optimal coercion cutoff \( \lambda_l \geq 1 \), which imposes a tighter condition than the feasibility constraint alone. That is, the liquidity shock must be large enough that the conglomerate cannot justify preserving any firm for long-term R\&D.

To determine when the interior solution yields such a threshold, recall the general formula for the interior coercion cutoff:
\begin{equation}
\lambda_l^{\text{int}} = \sqrt{ \frac{2\Delta L - X_1 \theta_1^s}{X_1 (\theta_1^l - \theta_1^s)} }
\end{equation}

The condition for this cutoff to reach or exceed 1 is:
\begin{align}
\lambda_l^{\text{int}} \geq 1 
\quad &\Leftrightarrow \quad 
\frac{2\Delta L - X_1 \theta_1^s}{X_1 (\theta_1^l - \theta_1^s)} \geq 1 \nonumber \\
&\Leftrightarrow \quad 
2\Delta L - X_1 \theta_1^s \geq X_1 (\theta_1^l - \theta_1^s) \nonumber \\
&\Leftrightarrow \quad 
2\Delta L \geq X_1 \theta_1^l \nonumber \\
&\Leftrightarrow \quad 
\Delta L \geq \frac{1}{2} X_1 \theta_1^l
\end{align}

Thus, the full coercion corner case arises not merely from feasibility (i.e., needing to liquidate some firms), but when the liquidity shock is severe enough to make preserving even the most promising firms suboptimal. This threshold, \( \Delta L \geq \frac{1}{2} X_1 \theta_1^l \), marks the boundary beyond which the interior allocation collapses into the full coercion regime.

\subsection{Cutoff Limits and Feasibility} \label{app:feasibility}
An interior solution arises only when the liquidity shock is below a critical threshold. 
\begin{align}
\lambda_l \leq 1
\end{align}
When the shock exceeds this threshold, the upper cutoff binds at $\lambda_l = 1$, and the conglomerate must rely solely on liquidation and coercion—no firms can be preserved on a long-term strategy.

Substitute the expression for \( \lambda_l \) in terms of \( \lambda_s \):
\begin{align}
\lambda_l = \frac{1+\Gamma}{\Gamma} \lambda_s \leq 1 \quad \Rightarrow \quad \lambda_s \leq \frac{\Gamma}{1 + \Gamma}
\end{align}

Now plug this upper bound for \( \lambda_s \) into the closed-form expression for \( \Delta L \) derived previously:
\begin{align}
\Delta L = \frac{1}{2} X_1 \lambda_s^2 \left[ (\theta_1^s - \theta_1^l) \left( \frac{1+\Gamma}{\Gamma} \right)^2 + (2\theta_1^l - \theta_1^s) \right]
\end{align}

Substitute \( \lambda_s = \frac{\Gamma}{1 + \Gamma} \) into the above expression:
\begin{align}
\lambda_s^2 = \left( \frac{\Gamma}{1 + \Gamma} \right)^2
\end{align}

Then:
\begin{align}
\Delta L &\leq \frac{1}{2} X_1 \left( \frac{\Gamma}{1 + \Gamma} \right)^2 \left[ (\theta_1^s - \theta_1^l) \left( \frac{1+\Gamma}{\Gamma} \right)^2 + (2\theta_1^l - \theta_1^s) \right] \nonumber\\
&= \frac{1}{2} X_1 \left[ (\theta_1^s - \theta_1^l) + (2\theta_1^l - \theta_1^s) \left( \frac{\Gamma}{1 + \Gamma} \right)^2 \right]
\end{align}

Thus, the upper bound on \( \Delta L \) ensuring an interior solution is:
\begin{equation}
\Delta L \leq \frac{X_1}{2} \left[ (\theta_1^s - \theta_1^l) + (2\theta_1^l - \theta_1^s)\left( \frac{\Gamma}{1 + \Gamma} \right)^2 \right]
\end{equation}

This bound separates the interior regime from the corner regime with no long-term preservation.

\subsection{Parameter Verification for Figure~\ref{fig:full_solution_set}} \label{app:figure}

The parameterization:
\[
X_1 = 1, \quad \theta_1^l = 0.3, \quad \theta_1^s = 0.7
\]
yields:
\[
\theta_1^s - \theta_1^l = 0.4, \quad 2\theta_1^l - \theta_1^s = -0.1
\]

For \( \theta_2^l = 1 \), \( \theta_2^s = 0.4 \), I compute:
\[
\Gamma = \frac{(\theta_2^l - \theta_2^s) \cdot \theta_1^l}{(\theta_1^s - \theta_1^l) \cdot \theta_2^l} 
= \frac{0.18}{0.4} = 0.45
\]

So the cutoff ratio is:
\[
\lambda_l = \frac{1 + \Gamma}{\Gamma} \lambda_s = \frac{1.45}{0.45} \lambda_s \approx 3.22 \lambda_s
\]

This implies a wide coercion margin under moderate shocks. As \( \Delta L \to 0.15 \), the coercion margin saturates and \( \lambda_l \to 1 \), consistent with the transition to the full coercion corner regime.

\bigskip

To verify the threshold for preservation explicitly, set:
\[
\lambda_l = 1 = \frac{1 + \Gamma}{\Gamma} \lambda_s \quad \Rightarrow \quad \lambda_s = \frac{\Gamma}{1 + \Gamma}
\]
With \( \Gamma = 0.45 \), this gives:
\[
\lambda_s = \frac{0.45}{1.45} \approx 0.3103
\]

Substituting into the liquidity constraint:
\[
\Delta L = \frac{1}{2} X_1 \left[ \theta_1^s - (\theta_1^s - \theta_1^l)\lambda_s^2 \right] 
= \frac{1}{2} \left[ 0.7 - 0.4 \cdot (0.3103)^2 \right] 
\approx \frac{1}{2} \left[ 0.7 - 0.0385 \right] \approx 0.3308
\]

Thus, a liquidity shock of \( \Delta L = 0.33 \) generates the exact transition to the full coercion regime. A slightly smaller shock (\( \Delta L = 0.2 \)) would generate full liquidation, i.e., \( \lambda_s = 1 \), verified by:
\[
\Delta L = \frac{1}{2} X_1 \theta_1^l = \frac{1}{2} \cdot 0.3 = 0.15
\]

This confirms that the parameterization is consistent with both corner thresholds: 
\( \Delta L \geq 0.15 \) for some liquidation,
\( \Delta L \geq 0.33 \) for no long-term investment (full coercion).


\section{Growth and Misallocation Derivations}
\label{appendix:growth_derivations}

\renewcommand{\theequation}{B.\arabic{equation}}
\setcounter{equation}{0}

This appendix provides detailed derivations of the aggregate growth path under the conglomerate’s optimal cutoff policy and under a conservative coercion-dominant strategy. I explicitly characterize the growth loss from misallocation and analyze its behavior as the economy is far from and is close to the technology frontier.

\subsection{Liquidation Margin in the Dynamic Model}
\label{appsub:liquidation_margin}

In the dynamic setting, the conglomerate adjusts its liquidation margin \( \lambda_s \) each period to satisfy an endogenously scaled liquidity constraint. The expression below shows how \( \lambda_s \) evolves with the size of the shock, the distance to the technology frontier, and the relative payoff structure of short- and long-term innovation.

\begin{align}
    \lambda_s 
    &= \sqrt{ \frac{2\Delta L_t}
    {X_1\left[(\theta_1^s - \theta_1^l)\left(\frac{1 + \Gamma}{\Gamma}\right)^2 + (2\theta_1^l - \theta_1^s)\right]}} \nonumber\\
    &= \sqrt{ \frac{2\Delta L_t}
    {\beta \left[(\overline{A}_{t-1} - A_{t-1})\left(\frac{1 + \Gamma}{\Gamma}\right)^2 + (2A_{t-1} - \overline{A}_{t-1})\right]}} \nonumber\\
    &= \sqrt{ \frac{2 \eta}
    {\beta  \left[(1-a_{t-1})\left(\frac{1 + \Gamma}{\Gamma}\right)^2 + (2a_{t-1} - 1)\right]}}
\end{align}

This formulation highlights how the liquidation margin increases with the size of the liquidity shock \( \eta \), and decreases as the economy nears the technology frontier \( a_{t-1} \to 1 \), provided the liquidity benefit of coercion also declines.

\subsection{Aggregate Growth under the Optimal Policy}
\label{appsub: optimal_growth}

Let $A_t(i)$ denote the productivity of firm $i$ at time $t$. Given the productivity evolution described in Section~\ref{sec:growth}, aggregate productivity growth is defined by:
\begin{align}
1 + g_t &= \int_0^1 \frac{A_t(i)}{A_{t-1}(i)} \, di.
\end{align}

Using the innovation payoff structure,
\begin{align}
A_t(i) = \lambda \theta_2^h X_2,
\end{align}

with $h = s$ for short-term and $h = l$ for long-term strategies, and normalizing $A_{t-1}$, I decompose growth into strategy regions:
\begin{align}
1 + g_t 
&= \frac{1}{A_{t-1}} \left[ \int_{\lambda_s}^{\lambda_l} \lambda \theta_2^s X_2 \, d\lambda + \int_{\lambda_l}^1 \lambda \theta_2^l X_2 \, d\lambda \right] \nonumber\\
&= \frac{X_2}{A_{t-1}} \left[ \theta_2^s \frac{\lambda_l^2 - \lambda_s^2}{2} + \theta_2^l \frac{1 - \lambda_l^2}{2} \right].
\end{align}

Substitute $X_2 = 1$, $\theta_2^s = \beta^s \overline{A}_{t-1}$, $\theta_2^l = \beta^l A_{t-1}$, and define $a_t = A_{t-1} / \overline{A}_{t-1}$ to get:
\begin{align}
1 + g^*(a) &= \frac{1}{2a} \left[ (\lambda_l^2 - \lambda_s^2) \beta^s + (1 - \lambda_l^2) \beta^l a \right].
\end{align}

\subsection{Growth under the Coercion Trap (No Liquidation)}
\label{appsub:coercion_growth}

Suppose the conglomerate adopts a conservative policy that avoids liquidation, setting \(\lambda_s = 0\). The entire liquidity burden is met by coercing firms into short-term strategies. The liquidity constraint becomes:
\begin{align}
\Delta L &= \int_0^{\lambda_l} (\theta_1^s - \theta_1^l)\lambda X_1 \, d\lambda 
= \frac{1}{2} (\theta_1^s - \theta_1^l) X_1 \lambda_l^2.
\end{align}

Solving for \(\lambda_l\), using \(\theta_1^s = \overline{A}_{t-1}\), \(\theta_1^l = A_{t-1}\), \(X_1 = \beta\), and \(\Delta L = \eta \overline{A}_{t-1}\):
\begin{align}
\lambda_l^c 
= \sqrt{ \frac{2 \eta \overline{A}_{t-1}}{\beta (\overline{A}_{t-1} - A_{t-1})} }
= \sqrt{ \frac{2 \eta}{\beta (1 - a_{t-1})} }.
\end{align}

Because \(\lambda \in [0, 1]\), I cap this cutoff at:
\begin{align}
\tilde{\lambda}_l^c = \min \left\{ 1, \lambda_l^c \right\}.
\end{align}

Growth under this policy is:
\begin{align}
1 + g_c(a) &= \frac{1}{2a} \left[ \tilde{\lambda}_l^{c\,2} \beta^s + (1 - \tilde{\lambda}_l^{c\,2}) \beta^l a \right].
\end{align}

\subsection{Limit Behavior: Detailed Derivations}

I now derive the limiting behavior of growth losses under misallocation as $a \to \beta$ (early development) and $a \to 1$ (advanced development). These limits help quantify how the relative performance of the optimal policy compares to misallocation regimes as the economy evolves.

\subsubsection{Coercion Trap: $\lambda_s = 0$ (No Liquidation)}
\label{appsub:coercion_trap_limit}

Recall that under the coercion trap, all liquidity pressure is resolved through short-term coercion. The resulting cutoff is:
\begin{align}
\lambda_l^c &= \sqrt{ \frac{2 \eta}{\beta (1 - a)} }, \\
\tilde{\lambda}_l^c &= \min\left\{ 1, \lambda_l^c \right\}, \\
1 + g_c(a) &= \frac{1}{2a} \left[ \tilde{\lambda}_l^{c\,2} \beta^s + (1 - \tilde{\lambda}_l^{c\,2}) \beta^l a \right].
\end{align}

The optimal growth rate is:
\begin{align}
1 + g^*(a) &= \frac{1}{2a} \left[ (\lambda_l^2 - \lambda_s^2) \beta^s + (1 - \lambda_l^2) \beta^l a \right],
\end{align}
where $\lambda_s$ and $\lambda_l$ depend on $\Gamma(a) = \frac{\beta^l a - \beta^s}{1 - a}$.

\paragraph{Limit as $a \to \beta$}

At the early stages of development, I take $a \to \beta$. Note that:
\begin{align}
\Gamma(a) &= \frac{\beta^l a - \beta^s}{1 - a} \to 0^+ \quad \text{as } a \to \beta, \\
\lambda_s^* &\to 0, \\
\lambda_l^* &= \left( 1 + \frac{1}{\Gamma} \right) \lambda_s^* \to 0.
\end{align}

Thus,
\begin{align}
g^*(a) &\to \frac{1}{2a} \left[ 0 + (1 - 0) \beta^l a \right] = \frac{1}{2} \beta^l.
\end{align}

Under the coercion trap,
\begin{align}
\lambda_l^c &\to \sqrt{ \frac{2 \eta}{\beta (1 - \beta)} }, \quad \text{a constant}, \\
\Rightarrow \tilde{\lambda}_l^c &= \min\left\{ 1, \sqrt{ \frac{2 \eta}{\beta (1 - \beta)} } \right\}, \\
\Rightarrow g_c(a) &\to \frac{1}{2 \beta} \left[ \tilde{\lambda}_l^{c2} \beta^s + (1 - \tilde{\lambda}_l^{c2}) \beta^l \beta \right].
\end{align}

Subtracting:
\begin{align}
\lim_{a \to \beta} \text{Loss}_c(a) &= \frac{1}{2} \beta^l - \frac{1}{2\beta} \left[ \tilde{\lambda}_l^{c2} \beta^s + (1 - \tilde{\lambda}_l^{c2}) \beta^l \beta \right] \nonumber\\
&= \frac{1}{2\beta} \left[ \beta \beta^l - \tilde{\lambda}_l^{c2} \beta^s - (1 - \tilde{\lambda}_l^{c2}) \beta^l \beta \right] \nonumber\\
&= \frac{1}{2\beta} \left[ \tilde{\lambda}_l^{c2} (\beta^l \beta - \beta^s) \right].
\end{align}

Since $\beta^s < \beta^l \beta$, the term is strictly positive, and I have:
\begin{align}
\boxed{
\lim_{a \to \beta} \text{Loss}_c(a) = \frac{1}{2\beta} \tilde{\lambda}_l^{c2} (\beta^l \beta - \beta^s) > 0.
}
\end{align}

\paragraph{Limit as $a \to 1$}

As the economy nears the frontier:
\begin{align}
\Gamma(a) &\to \infty, \quad \lambda_l^* \to \lambda_s^*, \nonumber\\
\Rightarrow g^*(a) &\to \frac{1}{2} \beta^l.
\end{align}

Under the coercion trap:
\begin{align}
\lambda_l^c &\to \infty, \quad \tilde{\lambda}_l^c \to 1, , \nonumber\\
\Rightarrow g_c(a) &\to \frac{1}{2a} \beta^s \to \frac{1}{2} \beta^s.
\end{align}

Therefore:
\begin{align}
\boxed{
\lim_{a \to 1} \text{Loss}_c(a) = \frac{1}{2}(\beta^l - \beta^s) > 0.
}
\end{align}

\subsubsection{Liquidation Fallacy: $\lambda_s = \lambda_l$ (No Coercion)}
\label{appsub:liquidation_fallacy_limit}

Under this policy, firms below $\lambda^\ell$ are liquidated:
\begin{align}
\Delta L &= \frac{1}{2} \overline{A}_{t-1} \beta (\lambda^\ell)^2 = \eta \overline{A}_{t-1}, \\
\Rightarrow \lambda^\ell &= \sqrt{ \frac{2 \eta}{\beta} }.
\end{align}

Since $\lambda_s = \lambda_l = \lambda^\ell$, the growth rate is:
\begin{align}
1 + g_\ell(a) &= \frac{1}{2a} (1 - (\lambda^\ell)^2) \beta^l a = \frac{1}{2}(1 - \lambda^{\ell 2}) \beta^l.
\end{align}

\paragraph{Limit as $a \to \beta$}

From the optimal policy:
\begin{align}
\lambda_s^* &\to 0, \quad \lambda_l^* \gg \lambda_s^*, \\
\Rightarrow g^*(a) &\to \frac{1}{2} \beta^l.
\end{align}

The liquidation policy is constant in $a$:
\begin{align}
g_\ell(a) &\to \frac{1}{2} \left( 1 - \frac{2 \eta}{\beta} \right) \beta^l \quad \text{if } \frac{2\eta}{\beta} < 1.
\end{align}

Hence:
\begin{align}
\boxed{
\lim_{a \to \beta} \text{Loss}_\ell(a) = \frac{1}{2} \left( \frac{2 \eta}{\beta} \right) \beta^l = \frac{\eta \beta^l}{\beta} > 0.
}
\end{align}

\paragraph{Limit as $a \to 1$}

As $a \to 1$, the optimal policy shifts fully to long-term strategies:
\begin{align}
\lambda_s^* \to 1, \quad g^*(a) \to \frac{1}{2} \beta^l.
\end{align}

The liquidation fallacy policy keeps $\lambda^\ell = \sqrt{ \frac{2 \eta}{\beta} }$, but:
\begin{align}
\tilde{\lambda}_s^\ell = \max\{ \lambda^\ell, \lambda_s^* \} \to 1, \\
\Rightarrow g_\ell(a) \to 0.
\end{align}

Therefore:
\begin{align}
\boxed{
\lim_{a \to 1} \text{Loss}_\ell(a) = \frac{1}{2} \beta^l > 0.
}
\end{align}

\medskip

Both misallocation regimes (coercion trap and liquidation fallacy) impose large growth losses in certain regimes of development. Coercion is especially harmful near the frontier, while liquidation imposes unnecessary sacrifice of viable firms early on.

\newpage
\bibliographystyle{apalike}
\bibliography{BF_RD}

\begin{thebibliography}{}

\bibitem[Acemoglu et~al., 2006]{Acemoglu2006DistanceTF}
Acemoglu, D., Aghion, P., and Zilibotti, F. (2006).
\newblock Distance to frontier, selection, and economic growth.
\newblock {\em Journal of the European Economic Association}, 4(1):37--74.

\bibitem[Acemoglu et~al., 2018]{acemoglu2018micro}
Acemoglu, D., Akcigit, U., Alp, H., Bloom, N., and Kerr, W. (2018).
\newblock Innovation, reallocation, and growth.
\newblock {\em American Economic Review}, 108(11):3450--91.

\bibitem[Acemoglu and Zilibotti, 1997]{acemoglu1997prometheus}
Acemoglu, D. and Zilibotti, F. (1997).
\newblock Was prometheus unbound by chance? risk, diversification, and growth.
\newblock {\em Journal of Political Economy}, 105(4):709--51.

\bibitem[Aghion et~al., 2010]{aghion2010volatility}
Aghion, P., Angeletos, G.-M., Banerjee, A., and Manova, K. (2010).
\newblock Volatility and growth: Credit constraints and the composition of
  investment.
\newblock {\em Journal of Monetary Economics}, 57(3):246--265.

\bibitem[Aghion and Howitt, 1992]{aghion1992model}
Aghion, P. and Howitt, P. (1992).
\newblock A model of growth through creative destruction.
\newblock {\em Econometrica}, 60(2):323--51.

\bibitem[Aghion and Howitt, 2005]{aghion2005quality}
Aghion, P. and Howitt, P. (2005).
\newblock Growth with quality-improving innovations: An integrated framework.
\newblock In Aghion, P. and Durlauf, S., editors, {\em Handbook of Economic
  Growth}, volume 1, Part A, chapter~02, pages 67--110. Elsevier, 1 edition.

\bibitem[Akcigit and Kerr, 2018]{akcigit2017growth}
Akcigit, U. and Kerr, W. (2018).
\newblock Growth through heterogeneous innovations.
\newblock {\em Journal of Political Economy}, 126(4):1374 -- 1443.

\bibitem[Allen and Gale, 2000]{allen2000comparing}
Allen, F. and Gale, D. (2000).
\newblock {\em Comparing Financial Systems}.
\newblock MIT Press, Cambridge, MA.

\bibitem[Anzoategui et~al., 2019]{anzoategui2019endogenous}
Anzoategui, D., Comin, D., Gertler, M., and Martinez, J. (2019).
\newblock Endogenous technology adoption and r\&d as sources of business cycle
  persistence.
\newblock {\em American Economic Journal: Macroeconomics}, 11(3):67--110.

\bibitem[Aoki and Patrick, 1995]{horiuchi1995japan}
Aoki, M. and Patrick, H., editors (1995).
\newblock {\em The Japanese Main Bank System: Its Relevance for Developing and
  Transforming Economies}.
\newblock Oxford University Press.

\bibitem[Barlevy, 2004]{barlevy2004timing}
Barlevy, G. (2004).
\newblock On the timing of innovation in stochastic schumpeterian growth
  models.
\newblock Technical Report WP-04-11, Federal Reserve Bank of Chicago.

\bibitem[Belenzon and Berkovitz, 2010]{belenzon2019innbusgroup}
Belenzon, S. and Berkovitz, T. (2010).
\newblock Innovation in business groups.
\newblock {\em Management Science}, 56(3):519--535.

\bibitem[Bernstein, 2022]{Bernstein2022TheEO}
Bernstein, S. (2022).
\newblock The effects of public and private equity markets on firm behavior.
\newblock {\em Annual Review of Financial Economics}, 14(1):295--318.

\bibitem[Bernstein et~al., 2019]{Bernstein2019PEfragile}
Bernstein, S., Lerner, J., and Mezzanotti, F. (2019).
\newblock Private equity and financial fragility during the crisis.
\newblock {\em The Review of Financial Studies}, 32(4):1309--1373.

\bibitem[Bianchi et~al., 2019]{bianchi2019growth}
Bianchi, F., Kung, H., and Morales, G. (2019).
\newblock Growth, slowdowns, and recoveries.
\newblock {\em Journal of Monetary Economics}, 101(C):47--63.

\bibitem[Bonciani et~al., 2023]{bonciani2023slow}
Bonciani, D., Gauthier, D., and Kanngiesser, D. (2023).
\newblock Slow recoveries, endogenous growth and macro-prudential policy.
\newblock {\em Review of Economic Dynamics}, 51:698--715.

\bibitem[Branstetter and Nakamura, 2003]{Branstetter2002HasJI}
Branstetter, L. and Nakamura, Y. (2003).
\newblock Is japan's innovative capacity in decline?
\newblock NBER Working Papers 9438, National Bureau of Economic Research, Inc.

\bibitem[Buera et~al., 2011]{buera2011financial}
Buera, F., Kaboski, J., and Shin, Y. (2011).
\newblock Finance and development: A tale of two sectors.
\newblock {\em American Economic Review}, 101(5):1964--2002.

\bibitem[Cerra et~al., 2021]{cerra2021financial}
Cerra, V., Hakamada, M., and Lama, R. (2021).
\newblock Financial crises, investment slumps, and slow recoveries.
\newblock Technical Report 2021/170, International Monetary Fund.

\bibitem[Chang et~al., 2001]{chang2001interpretkorean}
Chang, H.-J., Park, H.-J., and Yoo, C.~G. (2001).
\newblock Interpreting the korean crisis: Financial liberalization, industrial
  policy and corporate governance.
\newblock In Chang, H.-J., Palma, G., and Whittaker, D.~H., editors, {\em
  Financial Liberalization and the Asian Crisis}, chapter~9, pages 140--155.
  Palgrave Macmillan, London.

\bibitem[Deeg, 2001]{Deeg2001InstitutionalCA}
Deeg, R. (2001).
\newblock Institutional change and the uses and limits of path dependence: The
  case of german finance.
\newblock {\em ERN: Financial Markets}.

\bibitem[Demirguc-Kunt and Levine, 1999]{demirguckunt2001bankmarket}
Demirguc-Kunt, A. and Levine, R. (1999).
\newblock Bank-based and market-based financial systems - cross-country
  comparisons.
\newblock Policy Research Working Paper Series 2143, The World Bank.

\bibitem[Dewatripont and Maskin, 1995]{dewatripont1995credit}
Dewatripont, M. and Maskin, E. (1995).
\newblock Credit and efficiency in centralized and decentralized economies.
\newblock {\em The Review of Economic Studies}, 62(4):541--555.

\bibitem[Furman and Stiglitz, 1998]{furman1998eastasia}
Furman, J. and Stiglitz, J. (1998).
\newblock Economic crises: Evidence and insights from east asia.
\newblock {\em Brookings Papers on Economic Activity}, 29(2):1--136.

\bibitem[Gertler and Karadi, 2011]{gertler2011model}
Gertler, M. and Karadi, P. (2011).
\newblock A model of unconventional monetary policy.
\newblock {\em Journal of Monetary Economics}, 58(1):17--34.

\bibitem[Gertler and Kiyotaki, 2010]{gertler2010financial}
Gertler, M. and Kiyotaki, N. (2010).
\newblock Financial intermediation and credit policy in business cycle
  analysis.
\newblock In Friedman, B.~M. and Woodford, M., editors, {\em Handbook of
  Monetary Economics}, volume~3, chapter~11, pages 547--599. Elsevier, 1
  edition.

\bibitem[Gonenc et~al., 2007]{almeida2007internal}
Gonenc, H., Kan, O.~B., and Karadagli, E. (2007).
\newblock Business groups and internal capital markets.
\newblock {\em Emerging Markets Finance and Trade}, 43(2):63--81.

\bibitem[Gopalan et~al., 2007]{gopalan2007internal}
Gopalan, R., Nanda, V., and Seru, A. (2007).
\newblock Affiliated firms and financial support: Evidence from indian business
  groups.
\newblock {\em Journal of Financial Economics}, 86(3):759--795.

\bibitem[Gopalan et~al., 2014]{Goplan2014InternalCapDiv}
Gopalan, R., Nanda, V., and Seru, A. (2014).
\newblock Internal capital market and dividend policies: Evidence from business
  groups.
\newblock {\em The Review of Financial Studies}, 27(4):1102--1142.

\bibitem[Guerron and Jinnai, 2019]{guerronquintana2019financial}
Guerron, P. and Jinnai, R. (2019).
\newblock Financial frictions, trends, and the great recession.
\newblock {\em Quantitative Economics}, 10(2):735--773.

\bibitem[Harrigan, 2024]{harrigan2024rise}
Harrigan, K. (2024).
\newblock Rise of the new conglomerates.
\newblock Forthcoming. Columbia Business School Research.

\bibitem[Heshmati and Kim, 2011]{heshmati2011chaebol}
Heshmati, A. and Kim, H. (2011).
\newblock The r\&d and productivity relationship of korean listed firms.
\newblock {\em Journal of Productivity Analysis}, 36:125--142.

\bibitem[Hoshi and Kashyap, 2001]{hoshi2001corporate}
Hoshi, T. and Kashyap, A.~K. (2001).
\newblock {\em Corporate Financing and Governance in Japan: The Road to the
  Future}.
\newblock MIT Press, Cambridge, MA.

\bibitem[Joh, 2003]{joh2003corporate}
Joh, S.~W. (2003).
\newblock Corporate governance and firm profitability: evidence from korea
  before the economic crisis.
\newblock {\em Journal of Financial Economics}, 68(2):287--322.

\bibitem[Kaplan and Strömberg, 2008]{Kaplan2008LeveragedBA}
Kaplan, S.~N. and Strömberg, P. (2008).
\newblock Leveraged buyouts and private equity.
\newblock {\em The Review of Financial Studies}, 22(4).

\bibitem[Lamont, 1997]{Lamont1997CashFA}
Lamont, O. (1997).
\newblock Cash flow and investment: Evidence from internal capital markets.
\newblock {\em Journal of Finance}, 52(1):83--109.

\bibitem[Lin et~al., 2022]{lin2022distance}
Lin, J.~Y., Wang, W., and Xu, V.~Z. (2022).
\newblock Distance to frontier and optimal financial structure.
\newblock {\em Structural Change and Economic Dynamics}, 60(C):243--249.

\bibitem[Peek and Rosengren, 2005]{peek2005unnatural}
Peek, J. and Rosengren, E. (2005).
\newblock Unnatural selection: Perverse incentives and the misallocation of
  credit in japan.
\newblock {\em American Economic Review}, 95(4):1144--1166.

\bibitem[Queralto, 2020]{queralto2020slow}
Queralto, A. (2020).
\newblock A model of slow recoveries from financial crises.
\newblock {\em Journal of Monetary Economics}, 114(C):1--25.

\bibitem[Rajan, 1992]{rajan1992insiders}
Rajan, R. (1992).
\newblock Insiders and outsiders: The choice between informed and arm's-length
  debt.
\newblock {\em Journal of Finance}, 47(4):1367--400.

\bibitem[Rajan et~al., 2000]{Rajan2000CostDiv}
Rajan, R., Servaes, H., and Zingales, L. (2000).
\newblock The cost of diversity: The diversification discount and inefficient
  investment.
\newblock {\em Journal of Finance}, 55(1):35--80.

\bibitem[Rodrik, 2006]{rodrik2006goodbye}
Rodrik, D. (2006).
\newblock Goodbye washington consensus, hello washington confusion? a review of
  the world bank's economic growth in the 1990s: Learning from a decade of
  reform.
\newblock {\em Journal of Economic Literature}, 44(4):973--987.

\bibitem[Romer, 1990]{romer1990endogenous}
Romer, P. (1990).
\newblock Endogenous technological change.
\newblock {\em Journal of Political Economy}, 98(5):S71--102.

\bibitem[Scharfstein and Stein, 2000]{scharfstein2000theds}
Scharfstein, D. and Stein, J. (2000).
\newblock The dark side of internal capital markets: Divisional rent‐seeking
  and inefficient investment.
\newblock {\em Journal of Finance}, 55(6):2537--2564.

\bibitem[Shin and Stulz, 1998]{shin1998intcapeff}
Shin, H.-H. and Stulz, R. (1998).
\newblock Are internal capital markets efficient?
\newblock {\em The Quarterly Journal of Economics}, 113(2):531--552.

\bibitem[Stein, 1997]{stein1997internal}
Stein, J. (1997).
\newblock Internal capital markets and the competition for corporate resources.
\newblock {\em Journal of Finance}, 52(1):111--33.

\bibitem[Stiglitz, 2002]{stiglitz2002globalization}
Stiglitz, J.~E. (2002).
\newblock {\em Globalization and Its Discontents}.
\newblock W. W. Norton \& Company, New York.

\bibitem[von Thadden, 1995]{vonthadden1995long}
von Thadden, E.-L. (1995).
\newblock Long-term contracts, short-term investment and monitoring.
\newblock {\em The Review of Economic Studies}, 62(4):557--575.

\bibitem[Wu and Yao, 2012]{wu2012mainbank}
Wu, X. and Yao, J. (2012).
\newblock Understanding the rise and decline of the japanese main bank system:
  The changing effects of bank rent extraction.
\newblock {\em Journal of Banking \& Finance}, 36(1):36--50.

\end{thebibliography}

\end{document}